\documentclass[
11pt,%
final,%
tightenlines,%
notitlepage,%
twoside,%
onecolumn,%
nobibnotes,%
nofootinbib,%
superscriptaddress,%
noshowpacs,%
noshowydk,%
showdoi,%
centertags,%
maik]%
{revtex4n}

\makeatletter
\def\@bibstyle{agu}
\makeatother

\usepackage{bm}

\input jpack_e
\setcaptionmargin{5mm}
\onelinecaptionstrue

\columntovsizetrue
\allowdisplaybreaks

\begin{document}

\newtheorem{teo}{Theorem}
\newtheorem*{teon}{Theorem}
\newtheorem{lem}{Lemma}
\newtheorem*{lemn}{Lemma}
\newtheorem{prp}{Proposition}
\newtheorem*{prpn}{Proposition}
\newtheorem{ass}{Assertion}
\newtheorem*{assn}{Assertion}
\newtheorem{assum}{Assumption}
\newtheorem*{assumn}{Assumption}
\newtheorem{stat}{Statement}
\newtheorem*{statn}{Statement}
\newtheorem{cor}{Corollary}
\newtheorem*{corn}{Corollary}
\newtheorem{hyp}{Hypothesis}
\newtheorem*{hypn}{Hypothesis}
\newtheorem{con}{Conjecture}
\newtheorem*{conn}{Conjecture}
\newtheorem{dfn}{Definition}
\newtheorem*{dfnn}{Definition}
\newtheorem{problem}{Problem}
\newtheorem*{problemn}{Problem}
\newtheorem{notat}{Notation}
\newtheorem*{notatn}{Notation}
\newtheorem{quest}{Question}
\newtheorem*{questn}{Question}

\theorembodyfont{\rm}
\newtheorem{rem}{Remark}
\newtheorem*{remn}{Remark}
\newtheorem{exa}{Example}
\newtheorem*{exan}{Example}
\newtheorem{cas}{Case}
\newtheorem*{casn}{Case}
\newtheorem{claim}{Claim}
\newtheorem*{claimn}{Claim}
\newtheorem{com}{Comment}
\newtheorem*{comn}{Comment}

\theoremheaderfont{\it}
\theorembodyfont{\rm}

\newtheorem{proof}{Proof}
\newtheorem*{proofn}{Proof}

\selectlanguage{english}
\Rubrika{\relax}
\CRubrika{\relax}
\SubRubrika{\relax}
\CSubRubrika{\relax}
%

\def\JournalNumber{0}
\def\JournalVolume{00}
%
%
%
\nameVolumeRus{}
\CnameVolumeRus{}
\nameIssueRus{\No}
\CnameIssueRus{}
\namePartRus{}
\namePagesRus{}
\nameYearShortRus{}
\JournalNameRus{}
\TranslitJournalNameRus{}
\JournalName{Regular and Chaotic Dynamics}
\JournalISSNCode{1560-3547}
\IssuePrice{}
\TransYearOfIssue{0000}
\TransCopyrightYear{2016}%
\OrigYearOfIssue{}
\OrigCopyrightYear{2016}%
\OrigIssueNo{\JournalNumber}
\OrigVolumeNo{\JournalVolume}
\TransVolumeNo{\JournalVolume}
\TransIssueNo{\JournalNumber}
\TransPartNo{}
\SHORTjournalPREFIX{RCD} 
\LONGjournalPREFIX{RegDyn} 
\BatFileName{call make_ps.bat} 
\BatSwitch{3} 
\IssueName{}
\SupplementNumber{}
\PublicationSerialNumberInYear{0}
\PublicationSerialNumberInVolume{0}
\ConditionalIssueDate{"year","month","day","name","type"}
\PagePrefix{}
\JournalISSNonlineCode{}
\JournalISSNCodeRus{}
\JournalISSNonlineCodeRus{}
\VolumeName{}
\IssnoName{none}
\PartnoName{}
\FpageNamepp{}
\FpageNnamep{}
\FpagePrefix{}
\LpageNnamepp{}
\LpageNamep{}
\LpagePrefix{}
\VolumePageNumbering{}
\JournalPubID{}
\FirstJournalPageNumber{}
\LastJournalPageNumber{}
\makeatletter
\def\MAIKlogo{RCD Editorial Office}
\def\maikpraefix{10.0000/S}
\edef\@ContentsHeadLineB{Simultaneous English language translation of the journal is available from \noexpand\MAIKlogo}
\def\Distributed{Distributed worldwide by Springer. }
\def\ArticlePages#1{\relax}
\@ifxundefined\CONT@sw{\@booleantrue\CONT@sw}{}%
\@booleantrue\showPACS@sw%
\@booleantrue\showKEYS@sw %
\@booleantrue\noOrigJournalVersion@sw
\@booleantrue\noOrigVolumeNo@sw
\@booleanfalse\noTransVolumeNo@sw
\makeatother
\input maikdoi %

\beginpaper


\input engnames
\titlerunning{Lyapunov analysis of strange pseudohyperbolic attractors}
\authorrunning{Kuptsov and Kuznetsov}
\toctitle{Lyapunov analysis of strange pseudohyperbolic attractors}
\tocauthor{P.\,V.\,Kuptsov and S.\,P.\,Kuznetsov}
\title{Lyapunov analysis of strange pseudohyperbolic attractors:
angles between tangent subspaces, local volume expansion and
contraction}
\firstaffiliation{
}%
\articleinenglish 
\PublishedInRussianNo
\author{\firstname{Pavel\,V.}~\surname{Kuptsov}}%
\email[E-mail: ]{p.kuptsov@rambler.ru}
\affiliation{
Institute of electronics and mechanical engineering, Yuri
Gagarin State Technical University of Saratov\\
Politekhnicheskaya 77, Saratov 410054, Russia}%
\author{\firstname{Sergey\,P.}~\surname{Kuznetsov}}%
\email[E-mail: ]{spkuz@yandex.ru}
\affiliation{
Institute of Mathematics, Information Technologies and Physics,
Udmurt State University\\
Universitetskaya 1, Izhevsk 426034, Russia}%
\affiliation{
Kotel'nikov's Institute of Radio-Engineering and
Electronics of RAS, Saratov Branch\\
Zelenaya 38, Saratov 410019, Russia}
\begin{abstract}
  Pseudohyperbolic attractors are genuine str\-ange chaotic
  attractors. They do not contain stable periodic orbits and are
  robust in a sense that such orbits do not appear under
  variations. The tangent space of these attractors is split into a
  direct sum of volume expanding and contracting subspaces and these
  subspaces never have tangencies with each other. Any contraction in
  the first subspace, if occur, is weaker than contractions in the
  second one. In this paper we analyze local structure of several
  chaotic attractors recently suggested in literature as
  pseudohyperbolic. The absence of tangencies and thus the presence of
  the pseudohyperbolicity is verified using the method of angles that
  includes computation of distributions of the angles between the
  corresponding tangent subspaces. Also we analyze how volume
  expansion in the first subspace and the contraction in the second
  one occurs locally. For this purpose we introduce a family of
  instant Lyapunov exponents. Unlike the well known finite time ones,
  the instant Lyapunov exponents show expansion or contraction on
  infinitesimal time intervals. Two types of instant Lyapunov
  exponents are defined. One is related to ordinary finite time
  Lyapunov exponents computed in the course of standard algorithm for
  Lyapunov exponents. Their sums reveal instant volume expanding
  properties. The second type of instant Lyapunov exponents shows how
  covariant Lyapunov vectors grow or decay on infinitesimal
  time. Using both instant and finite time Lyapunov exponents we
  demonstrate that specific to the pseudohyperbolicity average
  expanding and contracting properties are typically violated on
  infinitesimal time. Instantly volumes from the first subspace can
  sometimes be contacted, directions in the second subspace can
  sometimes be expanded, and the instant contraction in the first
  subspace can sometimes be stronger than the contraction in the
  second subspace.
\end{abstract}
\received{September 1, 2018}
\accepted{Month XX, 20XX}%
\maketitle

\textmakefnmark{0}{)}%


\renewcommand\labelenumi{(\roman{enumi})}
\renewcommand\theenumi\labelenumi

\newcommand{\onefig}[1]{\centering{\includegraphics[width=0.6\columnwidth]{#1}}}

\newcommand{\mat}[1]{\bm{\mathrm{#1}}}

\newcommand{\clvmat}{\mat{\Gamma}}
\newcommand{\mydd}{\mathop{}\!\mathrm{d}}
\newcommand{\supT}{\text{T}}

\newcommand{\subbkw}{\text{b}}
\newcommand{\bkwmat}{\mat Q_\subbkw}
\newcommand{\bkwpre}{\widetilde{\mat Q}_\subbkw}
\newcommand{\bkwr}{\mat R_\subbkw}
\newcommand{\bkwa}{\mat A_\subbkw}

\newcommand{\bkwvec}{q}
\newcommand{\bkwmatelem}{q}

\newcommand{\subfwd}{\text{f}}
\newcommand{\fwdmat}{\mat Q_\subfwd}
\newcommand{\fwdpre}{\widetilde{\mat Q}_\subfwd}
\newcommand{\fwdr}{\mat R_\subfwd}
\newcommand{\fwda}{\mat A_\subfwd}

\newcommand{\fbas}{F}
\newcommand{\xbas}{X}

\newcommand{\xvar}{x}

\newcommand{\phdim}{N}
\newcommand{\nang}{K}

\newcommand{\fpropag}{\mat{\mathcal{F}}}

\newcommand{\ftledt}{\tau}
\newcommand{\numdt}{\Delta t}

\newcommand{\maxval}{\operatorname{max}}
\newcommand{\minval}{\operatorname{min}}
\newcommand{\clvec}{\gamma}

\newcommand{\jacob}{\mat{J}}
\newcommand{\jacobelem}{j}

\newcommand{\ible}{\Lambda}
\newcommand{\ftble}{\bar{\ible}}

\newcommand{\icle}{\mathcal{L}}
\newcommand{\ftcle}{\bar{\icle}}

\section*{INTRODUCTION}

Success in practical applications of chaotic theory essentially
depends on the robustness of the implemented systems. It means that
the chaotic regime must not be destroyed or qualitatively changed
under small variations of parameters of the
system~\cite{GenChaos}. Moreover, chaotic regime have to demonstrate
good stochastic properties proven by rigorous mathematical analysis.

One class satisfying these requirements contains systems with
uniformly hyperbolic chaos. Systems of this type, like, for example,
the Smale-Williams solenoid, manifest deterministic chaos justified in
rigorous mathematical sense. They demonstrate strong and structurally
stable stochastic properties~\cite{Smale67,Anosov95,KatHas95}. Though
many years hyperbolic attractors were considered only as a
mathematical abstraction, recently many examples of physically
realizable systems with hyperbolic attractor have been
suggested~\cite{HyperBook12,KuzUFN11}.

Uniformly hyperbolic attractors contain only saddle trajectories. For
discrete time systems these trajectories have well defined contracting
and expanding manifolds. The former contains phase trajectories
approaching the attractor in direct time and the latter corresponds to
the approaching in the reversed time. In the linear space of small
perturbations tangent to these manifolds this situation corresponds to
the splitting of the whole space into a direct sum of two subspaces
such that in one of them all directions are expanding and in the
second one they are contracting. The important feature of the saddle
trajectories and thus of the hyperbolic attractors is that the
contracting and expanding manifolds can intersect each other but can
not have tangencies. In the associated tangent space it is reflected
in the absence of clashes between vectors from the expanding and
contracting subspaces so that the angles between these subspaces never
vanish. For systems with continues time in addition to the expanding
and contracting tangent subspaces the neutral tangent subspace is
added, and all these three subspaces never have tangencies with each
other.

Besides the uniformly hyperbolic attractors one more class of systems
with a ``good'' chaos is formed by systems with pseudohyperbolic
attractors (the Lorenz attractor, ``wild''
attractors)~\cite{BeyondHyp,TurShil98,TurShil08,PsReview}. These
attractors are genuine strange attractors since each orbit has
positive Lyapunov exponent, i.e., stable periodic orbits are absent,
and this property is robust being preserved under at least small
perturbations. The tangent space of pseudohyperbolic systems is split
into a direct sum of volume expanding and contracting
subspaces. Notice that now only the expansion of volumes is required
instead of expansion along all direction needed for the uniform
hyperbolicity. These splitting must be invariant in time and the
subspaces can not have tangencies.

Necessary condition for the existence of the pseudohyperbolic
attractor is the following relation for its Lyapunov
exponents~\cite{TurShil98,TurShil08}:
\begin{equation}
  \label{eq:ps_necess_cond}
  \sum_{i=1}^n\lambda_i>0, \text{ and } \lambda_i<0 \text{ for } i>n.
\end{equation}
When the conditions~\eqref{eq:ps_necess_cond} hold, to confirm the
pseudohyperbolicity one also have to ensure that the $n$-dimensional
volume expanding subspace and $(\phdim-n)$-dimensional contracting
subspace do not have tangencies.

Based on discussions in
Refs.~\cite{TurShil98,TurShil08,PsReview,GonGonPshyp}, the following
list of properties of pseudohyperbolic attractors can be formulated:

\begin{enumerate}\itemsep=-2pt
\item \label{it:psh_ang} The tangent space is split into a direct sum
  of two hyperbolically isolated subspaces such that angles between
  them never vanish.
\item \label{it:psh_expand} The first $n$-dimensional subspace
  exponentially expands $n$-dimensional volumes, i.e., the sum of the
  Lyapunov exponents corresponding to this subspace is positive.
\item \label{it:psh_contr} The second subspace exponentially contracts
  all its vectors, i.e., all corresponding Lyapunov exponents are
  negative.
\item \label{it:psh_dosmmx} Any contraction in the first subspace, if
  occurs, is exponentially weaker than any contraction in the second
  subspace.
\end{enumerate}

In this paper we will test these properties for several concrete
examples of chaotic systems. The absence of the tangencies,
(property~\ref{it:psh_ang}) will be verified numerically using
suggested in Ref.~\cite{FastHyp12} implementation of the method of
angles. Three other properties are fulfilled automatically if the
necessary condition~\eqref{eq:ps_necess_cond} holds. However, unlike
the angles that are computed at the trajectory points with small step and
thus describe the attractor locally, Lyapunov exponents provide global
characteristics and ignore its fine details due to averaging. In this
paper we are going to test how the properties
\ref{it:psh_expand},~\ref{it:psh_contr}, and \ref{it:psh_dosmmx} are
fulfilled locally, on infinitesimal and short time intervals.

For this purpose, finite time Lyapunov exponents will be computed based
both on orthogonal Gram-Sch\-midt vectors and on covariant Lyapunov
vectors. Moreover instant Lyapunov exponents will be introduced that
provide expansion or contraction rates on infinitesimal time.

The paper is organized as follows. In Sect.~\ref{sec:theory} we will
briefly review the methods of computation of Lyapunov exponents,
covariant and orthogonal Lyapunov vectors, finite time Lyapunov
exponents. Also an instant Lyapunov exponents will be defined. The
main Sect.~\ref{sec:examp} is devoted to the testing of
pseudohyperbolicity of several attractors. Finally, in
Sect.~\ref{sec:conl} the results are discussed.

\section{\label{sec:theory}SOME BASICS OF LYAPUNOV ANALYSIS}

In this section we will briefly review methods of Lyapunov analysis
required for the further investigation of pseudohyperbolicity. We will
discuss the methods of computation of Lyapunov exponents, finite time
exponents, covariant Lyapunov vectors (CLVs) and angles between
tangent subspaces. Moreover we will introduce a family of instant
Lyapunov exponents that show the exponential growth rates in tangent
space on infinitesimal time.

\subsection{Covariant Lyapunov Vectors and Angles Between Tangent
  Subspaces}

Computation of angles between tangent subspaces can be done using
CLVs. These vectors are named ``covariant'' since $n$th vector at time
$t_1$ is mapped by a tangent flow to the $n$th vector at time $t_2$,
and a rate of its exponential expansion or contraction averaged over
an infinitely long trajectory is equal to the $n$th Lyapunov exponent
$\lambda_n$. Two algorithms for computation of these vectors were
first reported in the pioneering works~\cite{GinCLV,WolfCLV}. See also
paper~\cite{CLV2012} for more detailed explanation and one more
algorithm, and also a book~\cite{PikPol16} for a survey.

The importance of CLVs lies in the fact that they form a tangent basis
for expanding and contracting manifolds of trajectories of a dynamical
system. In particular, these vectors can indicate hyperbolicity of
chaos.  By the definition, both uniform hyperbolicity and its weaker
forms are related to the transversality of the tangent
subspaces~\cite{Smale67,Anosov95,KatHas95,Pesin04,BeyondHyp}. A
chaotic system is uniformly hyperbolic when expanding, contracting,
and also neutral, if any, subspaces are hyperbolically isolated, i.e.,
never have tangencies. In terms of CLVs it means that the angles
between the subspaces spanned by the corresponding CLVs never
vanish. In this paper we will put attention to the pseudohyperbolicity
which requires the absence of tangencies between volume expanding and
contracting subspaces~\cite{TurShil98,TurShil08,PsReview,GonGonPshyp}.

Verification of the hyperbolic isolation of tangent subspaces will be
done using the method of angles~\cite{FastHyp12} that in turn is based
on the method for CLVs computation suggested in Ref.~\cite{CLV2012} as
LU-method.

Consider a continues time system
\begin{equation}
  \label{eq:bas_equation}
  \dot\xbas(t)=\fbas(\xbas(t),t),
\end{equation}
where $\xbas\in\mathbb{R}^\phdim$ is $\phdim$-dimensional state
vector, and $\fbas$ is a nonlinear function. Infinitely small or
tangent perturbations to trajectories of the
system~\eqref{eq:bas_equation} obey the variational equation
\begin{equation}
  \label{eq:var_equation}
  \dot\xvar(t)=\jacob(t)\xvar(t),
\end{equation}
where $\xvar\in\mathbb{R}^\phdim$ is a tangent vector and
$\jacob(t)\in \mathbb{R}^{\phdim\times\phdim}$ is the Jacobian matrix,
i.e., the matrix of derivatives of $\fbas$ with respect to
$\xbas$. Its time dependence can be both implicit via $X(t)$ and
explicit (for non-autonomous case). For a discrete time system we
have:
\begin{align}
  \label{eq:bas_equation_discr}
  \xbas_{n+1}&=\fbas(\xbas_n,n),\\
  \label{eq:var_equation_discr}
  \xvar_{n+1}&=\jacob_n\xvar_n.
\end{align}
Here all terms have the same meaning as above, and $n$ denotes
discrete time.

Both for continues and discrete time systems the evolution of the
tangent vectors from time $t_1$ to time $t_2$ can be expressed as
follows:
\begin{equation}
  \label{eq:propag}
  \xvar(t_2)=\fpropag(t_1,t_2)\xvar(t_1),
\end{equation}
where $\fpropag(t_1,t_2)$ is a linear operator called propagator. For
discrete time systems this is merely $t_2-t_1$ times iterated Jacobian
matrix of the system, and for continues time system the propagator is
built from Jacobian matrix using the Magnus
expansion~\cite{CLV2012}. In numerical simulations the action of the
propagator $\fpropag(t_1,t_2)$ is equivalent to solving variational
equation~\eqref{eq:var_equation} or~\eqref{eq:var_equation_discr} from
$t_1$ to $t_2$ simultaneously with the basic
system~\eqref{eq:bas_equation} or~\eqref{eq:bas_equation_discr},
respectively.

Computation routines for Lyapunov exponents and CLVs use inner
products of tangent vectors. Its particular form can be chosen
arbitrary, and Lyapunov exponents as well as CLVs do not depend on
this choice. However, in some cases finding an appropriate form for
the inner product is important for clarifying the correspondence
between mathematical models and numerical approximations. For example
in Refs.~\cite{KupKuz16,HypManyDelay2018} a special form of the inner
product is introduced for analysis of hyperbolicity of chaos in time
delay systems. In our analysis however, it is enough to consider the
simplest standard dot product.

Discussed algorithms for CLVs and angles are based on the standard
algorithm for Lyapunov exponents created independently and
simultaneously by Benettin et al.~\cite{Benettin80} and by Shimada and
Nagashima~\cite{Shimada79}. Assume we need to compute $\nang$ Lyapunov
exponents, or CLVs, or going to evaluate first $\nang$ angles between
the tangent subspaces. First, we initialize a set of $\nang$ unit
random tangent vectors, orthogonal to each other, and gather them as
columns of a matrix $\bkwmat(t_1)$. Applying the propagator
$\fpropag(t_1,t_2)$ to this matrix we obtain a set of vectors
$\bkwpre(t_2)$, now non orthogonal. We recall that in practice it
merely means that we solve variational equations from $t_1$ to $t_2$
$\nang$ times (for each column of $\bkwmat(t_1)$). Now we need to
orthogonalize $\bkwpre(t_2)$. There are many algorithms to do it. The
most known is called Gram-Schmidt orthogonalization. In more general
form this procedure is referred to as QR factorization and consists in
representation of the matrix as a product of an orthogonal $\mat Q$
and an upper triangular $\mat R$
matrices~\cite{GolubLoan,Hogben2013}. Thus, one iteration of the
standard algorithm includes the following operations:
\begin{align}
  \label{eq:forward_lyap_step}
  &\fpropag(t_1,t_2)\bkwmat(t_1)=\bkwpre(t_2),\\
  \label{eq:forward_qr}
  &\bkwpre(t_2)=\bkwmat(t_2)\bkwr(t_1,t_2).
\end{align}
The orthogonal matrix $\bkwmat(t_2)$ is used for the next stage of the
algorithm.

After skipping some transient, we can consider logarithms of diagonal
elements of $\bkwr(t_1,t_2)$. Dividing them by the corresponding time
step, $\ftledt=t_2-t_1$, we obtain finite time Lyapunov exponents
(FTLEs) associated with the time interval $\ftledt$, and averaging
them over long trajectory we obtain numerical approximations for
global Lyapunov exponents $\lambda_i$. In what follows, $\lambda_i$
will be referred to as merely Lyapunov exponents.

The algorithm for CLVs and angles that we use here requires the matrix
$\bkwmat(t)$. After the transient, the columns of this orthogonal
matrix turns to the backward Lyapunov vectors. This name seems to be
counterintuitive, but its origin is not related to the direction of
iterations in time. It indicates that they have arrived at the current
point after long evolution initialized in the far
past~\cite{AGuide,CLV2012}. Also these vectors are known as
Gram-Schmidt vectors. The directions pointed by these vectors, except
the first one, depend on the choice of the inner product. It means
that individually they do not bring much information about the tangent
space structure. But the subspaces they span, do. Assume that we
already have found CLVs and they are gathered as columns of the matrix
$\clvmat(t)$. The backward Lyapunov vectors form an orthogonal matrix
in QR-decomposition of $\clvmat(t)$~\cite{CLV2012}:
\begin{equation}
  \label{eq:clv_via_bkw}
  \clvmat(t)=\bkwmat(t)\bkwa(t),
\end{equation}
where $\bkwa(t)$ is an upper triangular matrix. Since QR-decomposition
preserves subspaces spanned by vector-columns of the decomposed
matrix (see, for example book~\cite{GolubLoan} for details),
Eq.~\eqref{eq:clv_via_bkw} shows that the first CLV coincides with the
first backward Lyapunov vector, the second one lies in a plane spanned
by the first two backward vectors, the third one belongs to a
three-dimensional space of the first three backward vectors and so on.

The second part of the discussed algorithm includes iterations with
the adjoint propagator. Notice that the action of the adjoint
propagator as well as the action of the inverted one corresponds to
steps backward in time~\cite{CLV2012}. The form of the adjoint
propagator depends on the chosen inner
product~\cite{KupKuz16,HypManyDelay2018}, and the standard dot product
produces its simplest version: the adjoint propagator is obtained from
the original one simply by transposition as
$\fpropag^\supT(t_1,t_2)$. The steps are performed again with $\nang$
vectors that are QR-decomposed after each action of the propagator
$\fpropag^\supT(t_1,t_2)$:
\begin{align}
  \label{eq:bkwadj_lyap_step}
  &\fpropag(t_1,t_2)^\supT\fwdmat(t_2)=\fwdpre(t_1),\\
  \label{eq:bkwadj_qr}
  &\fwdpre(t_1)=\fwdmat(t_1)\fwdr(t_1,t_2).
\end{align}
Here $\fwdmat(t)$ is an orthogonal matrix with $\nang$ columns. When
one drops out some transient, columns of $\fwdmat(t)$ becomes the so
called forward Lyapunov vectors. ``Forward'' here indicates that the
vectors arrive from far future.

Assume for a moment that we have the full set of $\phdim$ forward
vectors. Then the matrix $\fwdmat(t)$ is an orthogonal matrix in
QL-decomposition of the CLVs matrix $\clvmat(t)$:
\begin{equation}
  \label{eq:clv_via_fwd}
  \clvmat(t)=\fwdmat(t)\fwda(t).
\end{equation}
Here $\fwda(t)$ is a lower triangular matrix~\cite{CLV2012}. Thus the
$\phdim$th forward vector coincides with the last CLV, the last two
forward vectors span the subspace containing the $(\phdim-1)$th CLV,
and so on. It means that the remaining forward vectors, i.e., the
columns of $\fwdmat(t)$ from the $1$st to $n$th, form an orthogonal
complement for the subspace containing last $N-n$ CLVs. Thus, given
$\nang$ backward Lyapunov vectors in $\bkwmat(t)$ and $\nang$ forward
Lyapunov vectors in $\fwdmat(t)$, we have a subspace with the first
$\nang$ CLVs and an orthogonal complement for the subspace for
$\phdim-\nang$ remaining CLVs. It is enough to compute $\nang$ CLVs
and a series of angles between the subspaces spanned by these vectors.

Equating the left hand sides of Eqs.~\eqref{eq:clv_via_bkw}
and~\eqref{eq:clv_via_fwd} we obtain:
\begin{align}
  \label{eq:matrix_p}
  &\mat P(t)=[\fwdmat(t)]^\supT\bkwmat(t),\\
  \label{eq:clv_lu}
  &\mat P(t) \bkwa(t)=\fwda(t).
\end{align}
Thus, given $\bkwmat(t)$ and $\fwdmat(t)$, we first compute $\mat P(t)$
with Eq.~\eqref{eq:matrix_p}. Then, since $\bkwa(t)$ and $\fwda(t)$ are
upper and lower triangular matrices, respectively, they are computed
for $\mat P(t)$ form Eq.~\eqref{eq:clv_lu} as its LU decomposition,
see Ref.~\cite{CLV2012} for more details. Finally, using $\bkwa(t)$
and $\bkwmat(t)$ we can find CLVs from Eq.~\eqref{eq:clv_via_bkw}.

Angles between subspaces are called principal angles. Cosines of these
angles can be found as singular values of a matrix whose elements are
pairwise inner products of orthogonal base vectors for these
subspaces~\cite{GolubLoan}.  We have an orthogonal basis for the first
subspace of interest in $\bkwmat(t)$, also there is a basis for the
orthogonal complement of the second subspace in $\fwdmat(t)$, and
$\mat P(t)$ is the matrix of their inner products. Two $n$-dimensional
subspaces have $n$ principal angles. But since we are interested in
verification of tangencies of these subspaces we need only one of the
angles. Because $\fwdmat(t)$ is the orthogonal complement to the
subspace of interest, the tangency is signaled by the largest
principal angle that corresponds to the smallest singular value. Once
the matrix $\mat P$ is computed we can evaluate a series of $\nang$
angles. Taking top left square submatrices
$\mat P[1\colon n,1\colon n]$, where $n=1,2,\ldots,\nang$, and finding
their smallest singular values $\sigma_n$, we obtain the angle between
the $n$-dimensional subspace of the first CLVs and the
$(\phdim-n)$-dimensional subspace of the remaining CLVs as:
\begin{equation}
  \label{eq:the_angle}
  \theta_n=\pi/2-\arccos\sigma_n.
\end{equation}

The smallest singular value $\sigma_n$ as well as the angle $\theta_n$
vanishes when a tangency between the corresponding subspaces occurs.
Because trajectories with the exact tangencies are rather untypical,
in actual computations we register a tangency between subspaces if the
corresponding angle can be arbitrarily small.

\subsection{Finite Time Lyapunov Exponents}

Finite time Lyapunov exponents (FTLEs) characterize expansions and
contractions in phase space on finite time intervals. They are
obtained from logarithms of diagonal elements of the upper triangular
matrix $\bkwr(t_1,t_2)$ computed after each QR decomposition in the
course of computation of Lyapunov exponents, see
Eq.~\eqref{eq:forward_qr}:
\begin{equation}
  \label{eq:ftle_via_r}
  \ftble_n(t_1,t_2)=\frac{\log r_{nn}(t_1,t_2)}{t_2-t_1}.
\end{equation}
The Lyapunov exponents $\lambda_n$ are the averagings of FTLEs
$\ftble_n(t_1,t_2)$ over infinitely long trajectory. They always
appear in a descending order in computations and show an hierarchy of
expansions and contractions in the phase space.

Individual meaning of FTLEs~\eqref{eq:ftle_via_r}, except the first
one, is not so clear.  The first FTLE shows how a typical tangent
vector exponentially grows from $t_1$ to $t_2$. By construction, the
second FTLE is the rate of exponential growth along a direction
perpendicular to the fastest one. It has no much of physical meaning
by itself. However, the sum $\ftble_1(t_1,t_2)+\ftble_2(t_1,t_2)$
shows the exponential grows rate of a typical two-dimensional
area. Similarly, the third FTLE $\ftble_3(t_1,t_2)$ admits clear
interpretation being summed with two previous ones: this sum indicates
the rate of exponential growth of a typical three-dimensional
volume. The sum of the first $n$ FTLEs is a growth rate for a typical
$n$-dimensional volume in the tangent space.

When CLVs have became available due to the effective algorithms for
their computations~\cite{GinCLV,WolfCLV}, in addition to
FTLEs~\eqref{eq:ftle_via_r}, a new sort of finite time Lyapunov
exponents were introduced, computed as rates of exponential grows of
CLVs on time interval $t_2-t_1$, see Ref.~\cite{CLV2012}. We will
refer to them as FTCLE and denote as $\ftcle_n(t_1,t_2)$. Similar to
FTLEs~\eqref{eq:ftle_via_r}, these CLV based exponents also converge
to Lyapunov exponents on large times, but their meaning is
different. Each FTCLE characterizes an exponential expansion or
contraction rate along a covariant direction where on average the
expansion or contraction occurs according to the respective Lyapunov
exponent. Since this covariant directions pointed by CLVs are not
orthogonal, the sums of FTCLEs are not related to the rates of volumes
expansion or contractions.

In brief, FTLEs are based on backward Lyapunov vectors and are
appropriate for testing volume expanding properties in tangent
space. For this purpose they have to be summed, while individual
values of FTLEs except the first one have no much sense. FTCLEs are
based on covariant Lyapunov vectors and are good for testing tangent
vectors expansion or contraction. Their sums have no sense and one has
to consider their values individually.

The specific feature of both FTLEs and FTCLEs is that they are
computed for finite time intervals. One of the appropriate ways of
employing them is analysis of their fluctuations on asymptotically
large time intervals~\cite{StatMechLyap11}. However when local
properties are required, it is usually unclear which interval
$t_2-t_1$ is sufficiently small to give a representative
picture. Obviously this problem makes sense only for continues time
systems, while for discrete time systems the local properties are
recovered by FTLEs and FTCLEs computed for unit time steps
$t_2-t_1=1$.

\subsection{Instant Lyapunov Exponents}

To analyze tangent space expansion on infinitesimal time we will
introduce here the \emph{instant Lyapunov exponents}. Let us start
with the instant Lyapunov exponents based on backward Lyapunov vectors
that will be called IBLE and denoted as $\ible_i(t)$. They have to be
related to FTLEs $\ftble_i(t_1,t_2)$ as follows:
\begin{equation}
  \label{eq:ftble_integral}
  \ftble_i(t_1,t_2)=\frac{1}{t_2-t_1}
  \int_{t_1}^{t_2}\ible_i(t)\mydd t.
\end{equation}
On the other hand, by the definition, the sum of $n$ first FTLEs is an
exponential growth rate of $n$-dimensional volume:
\begin{equation}
  \sum_{i=1}^{n}\ftble_i(t_1,t_2)=\frac{1}{t_2-t_1}
  \log\frac{\mathrm{Vol}_n(t_2)}{\mathrm{Vol}_n(t_1)}.
\end{equation}
Substituting here Eq.~\eqref{eq:ftble_integral} and differentiating by
$t_2$, we obtain:
\begin{equation}
  \sum_{i=1}^{n}\ible_i(t_2)=
  \frac{\mydd}{\mydd t_2}\log \mathrm{Vol}_n(t_2).
\end{equation}
Here we took into account that $\mathrm{Vol}_n(t_1)$ does not depend
on $t_2$. Volume $\mathrm{Vol}_n(t_2)$ is equal to the product of $n$
first diagonal elements $r_{ii}$ of the upper triangular matrix
obtained after QR decomposition, see Eq.~\eqref{eq:forward_qr}. The
detailed explanation of it can be found in Ref.~\cite{CLV2012}. Hence
\begin{align}
  &\sum_{i=1}^{n}\ible_i(t_2)=
  \frac{\mydd}{\mydd t_2} \sum_{i=1}^{n} \log r_{ii},\\
  \label{eq:ible_almost_done}
  &\ible_i(t)=\dot r_{ii}/r_{ii}.
\end{align}

To proceed, consider a variational equation in the matrix form:
\begin{equation}
  \dot{\mat V}=\jacob\mat V,
\end{equation}
where $\mat V$ is a matrix of tangent vectors.  Substituting $\mat V$
with its QR-decomposition we obtain:
\begin{equation}
  \dot{\mat Q} \mat R + \mat Q \dot{\mat R}=\jacob\mat Q\mat R,
\end{equation}
or, after simple matrix algebra:
\begin{equation}
  \label{eq:dot_r_over_r}
  \dot{\mat R}\mat R^{-1}=\mat Q^\supT\jacob\mat Q-
  \mat Q^\supT\dot{\mat Q}.
\end{equation}
For any orthogonal time dependent matrix $\mat Q$ the product
$\mat Q^\supT\dot{\mat Q}$ is always skew-symmetric. It can be easily
checked by differentiation of the identity $\mat Q^\supT \mat Q=1$. It
means that the diagonal of $\mat Q^\supT\dot{\mat Q}$ contains only
zeros. Thus, substituting diagonal elements of the matrices from
Eq.~\eqref{eq:dot_r_over_r} to Eq.~\eqref{eq:ible_almost_done} we
obtain:
\begin{equation}
  \label{eq:ible}
  \ible_i(t)=\bkwvec_i^\supT(t)\jacob(t) \bkwvec_i(t),
\end{equation}
where $\bkwvec_i$ is $i$th backward Lyapunov vector. Thus, to compute
IBLE $\ible_i(t)$ in the course of usual routine for Lyapunov exponent
after steps~\eqref{eq:forward_lyap_step},~\eqref{eq:forward_qr} we
need to multiply each backward vector by the Jacobian matrix and then
to find the inner product with the vector itself.

Divergence of the vector field produced by the continues time system
\eqref{eq:bas_equation} is known to be equal to the instant
exponential volume contraction rate in the whole $\phdim$-dimensional
phase space~\cite{KatHas95}. It means that the sum of $\phdim$ IBLEs
have to be equal to this divergence. This is indeed the case. By the
definition, the divergence is equal to the sum of diagonal elements of
the Jacobian matrix. Thus
\begin{equation}
  \label{eq:sum_div}
  \begin{split}
    \sum_{i=1}^\phdim \ible_i&=
    \sum_{ijk}\bkwmatelem_{ij}\jacobelem_{ik}\bkwmatelem_{kj}=
    \sum_{ik}\jacobelem_{ik}\sum_{j}\bkwmatelem_{ij}\bkwmatelem_{kj}\\&=
    \sum_{ik}\jacobelem_{ik}\delta_{ik}=
    \sum_{i}\jacobelem_{ii}=\operatorname{div}\fbas,
  \end{split}
\end{equation}
where $\bkwmatelem_{ij}$ and $\jacobelem_{ik}$ are elements of
matrices $\bkwmat$ and $\jacob$, respectively, and $\delta_{ik}$ is
Kronecker's symbol. Notice that these calculations use merely the
orthogonality of $\bkwmat$, and do not employ its specific form. It
means that this equality is rather trivial and can not be used, for
example, for testing correctness of computations of IBLEs. Also notice
that similar equality for the corresponding FTLEs can be fulfilled
only approximately, since $\operatorname{div}\fbas$ is an instant
value and FTLEs are always related to a finite time interval.

Let us now turn to the finite time exponents based on CLVs. We will
call them ICLE, denote as $\icle_i(t)$, and introduce via the
following integral:
\begin{equation}
  \label{eq:ftcle_integral}
  \ftcle_i(t_1,t_2)=\frac{1}{t_2-t_1}\int_{t_1}^{t_2}\icle_i(t)\mydd t,
\end{equation}
where $\ftcle_i(t_1,t_2)$ are FTCLE, i.e, the mentioned above finite
time exponent based on CLVs. FTCLE is equal to the exponential growth
rate of $i$th CLV $\clvec_i(t)$ on the time interval
$(t_2-t_1)$\footnote{Notice that using this equation for straightforward
  computation of FTCLE, i.e., solving numerically variational equation
  with $\clvec_i(t_1)$ as an initial condition, one have to take a
  sufficiently short interval $t_2-t_1$. Though formally CLVs are
  preserved in the course of running along a trajectory, they are
  fragile in a sense that any error grows. Thus the numerical
  approximations of CLVs slowly diverge from their true directions and
  tend to align along the first CLV.}:
\begin{equation}
  \label{eq:ftcle}
  \ftcle_i(t_1,t_2)=\frac{1}{t_2-t_1}\log\left(\frac{\|\clvec_i(t_2)\|}
    {\|\clvec_i(t_1)\|}\right).
\end{equation}
Combining Eqs.~\eqref{eq:ftcle_integral} and \eqref{eq:ftcle} and
differentiating by $t_2$ we obtain:
\begin{equation}
  \label{eq:icle_derive1}
  \icle(t_2)=
  \frac{\mydd}{\mydd t_2}\log \|\clvec(t_2)\|.
\end{equation}
Here we took into account that $\|\clvec(t_1)\|$ does not depend on
$t_2$. Now we proceed as follows:
\begin{equation}
  \label{eq:icle_derive2}
  \begin{split}
    \icle(t_2)&=\frac{1}{2}\frac{\mydd}{\mydd t_2}\log
    \|\clvec(t_2)\|^2=
    \frac{1}{2\|\clvec(t_2)\|^2}\frac{\mydd}{\mydd t_2}
    \clvec(t_2)^\supT\clvec(t_2)\\&=
    \frac{1}{2\|\clvec(t_2)\|^2}\left[\dot{\clvec(t_2)}^\supT\clvec(t_2)+
      \clvec(t_2)^\supT\dot{\clvec(t_2)}\right].
  \end{split}
\end{equation}
By the defenetion, CLVs evolve according to the variational
equation~\eqref{eq:var_equation}. Thus:
\begin{equation}
    \icle(t_2)=\frac{1}{2\|\clvec(t_2)\|^2} \left\{
      \clvec(t_2)^\supT[\jacob(t_2)^\supT+\jacob(t_2)]\clvec(t_2)
    \right\}.
\end{equation}
Taking into account that CLVs are always computed with unit norms, we
obtain the final equation for ICLE:
\begin{equation}
  \label{eq:icle}
  \icle(t)=\frac{1}{2} \left\{
    \clvec(t)^\supT[\jacob(t)^\supT+\jacob(t)]\clvec(t)\right\}.
\end{equation}

Altogether, we deal with characteristic exponents of four
types. Summed FTLEs $\ftble_i$ and IBLEs $\ible_i$ indicate the volume
expansion occurring on finite time intervals and instantly,
respectively. FTLEs are related to IBLEs via
Eq.~\eqref{eq:ftble_integral}. FTCLEs $\ftcle_i$ and ICLEs $\icle_i$
show expansion along covariant directions on finite time intervals and
instantly, respectively. These are related with each other according
to Eq.~\eqref{eq:ftcle_integral}. In what follows, we will use all of
them to analyze structure of chaotic attractors.

\section{\label{sec:examp}VERIFICATION OF PSEUDOHYPERBOLICITY}

\subsection{Lorenz System}
We start with the famous Lorenz
system~\cite{Lorenz63,Sparrow82,SchusJust05}:
\begin{equation}\label{eq:lorenz}
  \begin{aligned}
    \dot x &= \sigma(y-x),\\
    \dot y &= x(r-z)  - y,\\
    \dot z &= xy - bz.
  \end{aligned}
\end{equation}
Parameters are $r=28$, $\sigma=10$, $b=8/3$. To solve numerically
these and other equatios we use the Runge-Kutta method of the fourth
order.

First of all we need Lyapunov exponents. For this purpose we will
solve Eq.~\eqref{eq:lorenz} simultaneously with its variational
equations with time step
$\numdt=10^{-4}$. Iterations~\eqref{eq:forward_lyap_step},~\eqref{eq:forward_qr}
are repeated until the maximal absolute error of $\lambda_i$ becomes
less than $\epsilon=10^{-5}$. These computations are repeated ten
times, and the resulting exponents are averaged. The results are
$\lambda_1=0.906$, $\lambda_2\approx 10^{-5}$, and
$\lambda_3=-14.573$. The second exponent must actually be put to zero
since it corresponds to the symmetry of the
equations~\eqref{eq:lorenz} with respect to time shifts. The values
agree well with the values reported in the literature, see for
example~\cite{FroyAlf84,KuzDynChaos06,ElegantChaos}.

The Lorenz system is known to be
pseudohyperbolic~\cite{TurShil98,PsReview,BeyondHyp,BykShil92}. Our
purpose here is to confirm this by testing the absence of tangencies
between the volume expanding and contracting subspaces according to
the property~\ref{it:psh_ang} formulated in the Introduction. Also we
will test how the properties~\ref{it:psh_expand},~\ref{it:psh_contr},
and \ref{it:psh_dosmmx} are fulfilled locally.

\begin{figure}[tbp]
  \onefig{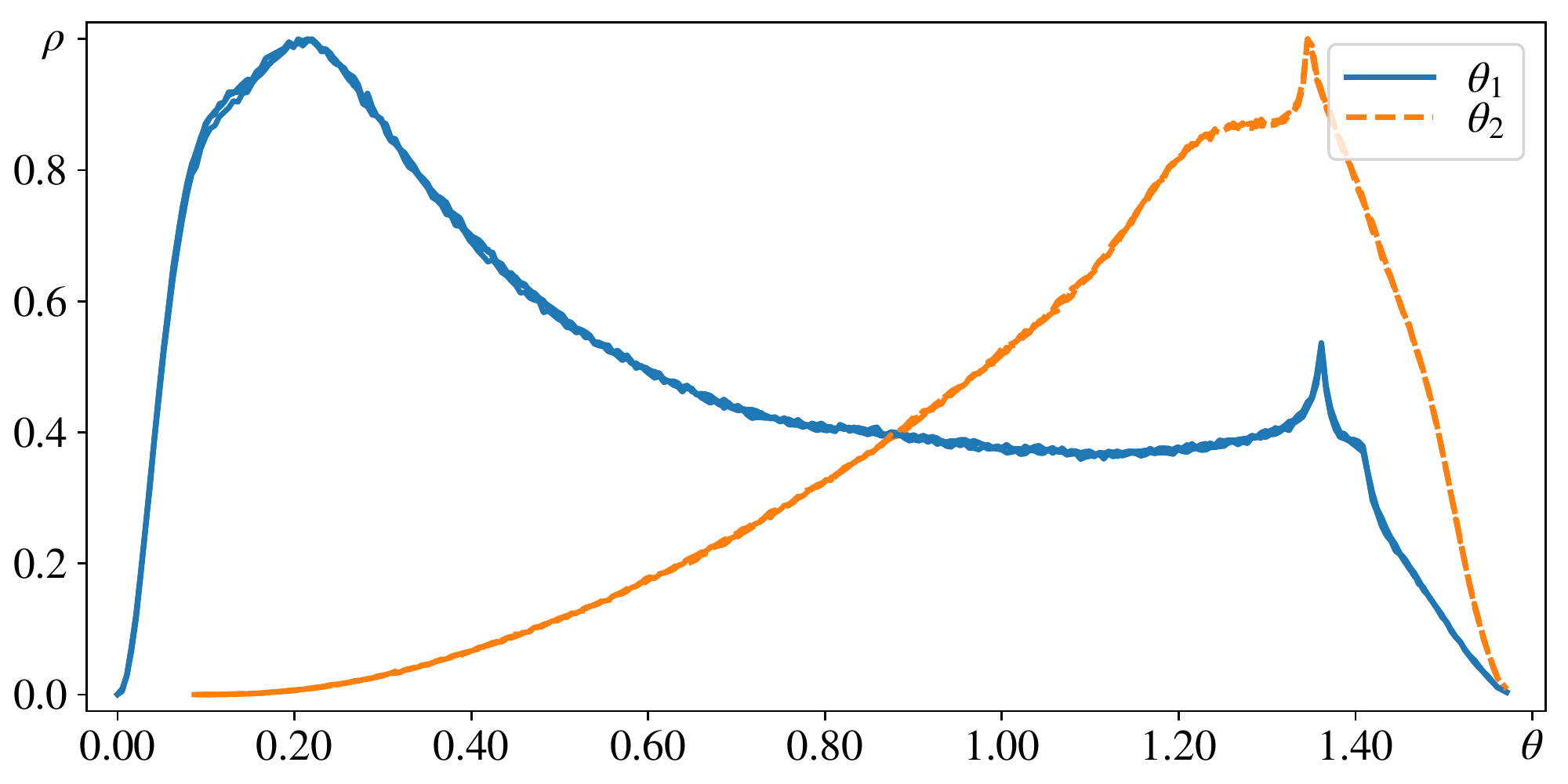}
  \caption{\label{fig:lorenz_ang}Distributions of angles between
    tangent subspaces for the Lorenz system~\eqref{eq:lorenz}. Each
    curve is computed three times with different numerical steps
    $\numdt=0.01$, $0.001$ and $0.0001$.  The curves almost perfectly
    coincide indicating that they are not affected by time step. The
    pseudohyperbolicity is confirmed by the non-vanishing $\theta_2$.}
\end{figure}

Since $\lambda_1+\lambda_2>0$ and $\lambda_3<0$, the tangent space of
the Lorenz system~\eqref{eq:lorenz} is expected to be split into
two-dimensional volume expanding and one-dimensional contracting
subspaces. The transversality of these two subspaces
(property~\ref{it:psh_ang}) is confirmed by Fig.~\ref{fig:lorenz_ang}
where distributions of angles between tangent subspaces are
shown. This and all subsequent figures have been plotted using
Matplotlib graphics package~\cite{Hunter:2007}. Angle $\theta_1$ is
computed between the subspace related to the first covariant vector
and the subspace spanned on two last ones; and $\theta_2$ is computed
between the subspace of the first two covariant vectors and the last
one. To check that the curves are not affected by the numerical step
size we have computed the angles three times, with steps
$\numdt=0.01$, $0.001$ and $0.0001$. (For the first curve the
orthogonalization and computation of the angles is done at each step,
for the second one after each $10$ steps and for the last one after
each $100$ steps.) As a result, all three curves are almost perfectly
coincide so that they are barely distinguishable in the figure. One
can see from the figure that the subspace of the two first vectors
never have common elements with the subspace of the last one, since
$\theta_2$ never vanishes. In other words these subspaces are
hyperbolically isolated. This is the main manifestation of the
pseudohyperbolicity. (Notice that the uniform hyperbolicity requires
the separation of the expanding, neutral and contracting subspaces,
i.e, those spanned by covariant vectors associated with positive, zero
and negative Lyapunov exponents. In particular in
Fig.~\ref{fig:lorenz_ang} the angle $\theta_1$ would also be nonzero.)

\begin{figure}[tbp]
  \onefig{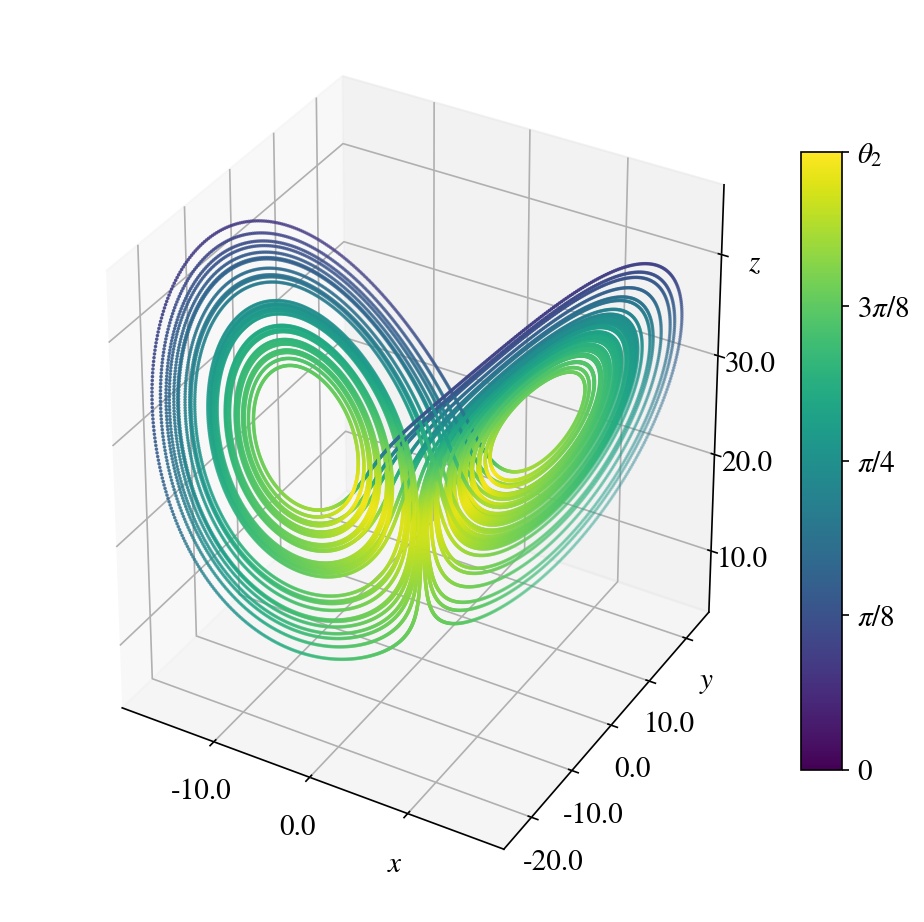}
  \caption{\label{fig:lorenz_attr3d}Phase portrait of the Lorenz
    system~\eqref{eq:lorenz}. Point colors correspond to values of the
    angle $\theta_2$ between the first two-dimensional tangent
    subspace and the second one-dimensional one. Observe that the
    small angles are located on attractor edges.}
\end{figure}

Figure~\ref{fig:lorenz_attr3d} shows the phase portrait of the Lorenz
system where the attractor points are painted according to the values
of $\theta_2$: lighter colors represent larger angles and darker
correspond to smaller ones. One can see that the large angles can be
found in inner areas of the attractor, while the most small values
(but nevertheless nonzero as indicates Fig.~\ref{fig:lorenz_ang}) are
located on its edges.

Let us now consider the property~\ref{it:psh_expand}, concerning the
volume expansion. As discussed, the volume expansion properties can be
tested using Lyapunov exponents corresponding to backward Lyapunov
vectors. Instant and finite time expansion of $n$-dimensional volumes
are characterized by $n$ summed IBLEs $\ible_i$ and FTLEs $\ftble_i$,
respectively:
\begin{equation}
  \label{eq:ible_acm}
  S_n(t)=\sum_{i=1}^{n}\ible_i(t),\;\;
  \bar{S}_n(t,t+\ftledt)=\sum_{i=1}^{n}\ftble_i(t,t+\ftledt).
\end{equation}
We recall that $\ftledt$ denotes here the averaging time.

\begin{figure}[tbp]
  \onefig{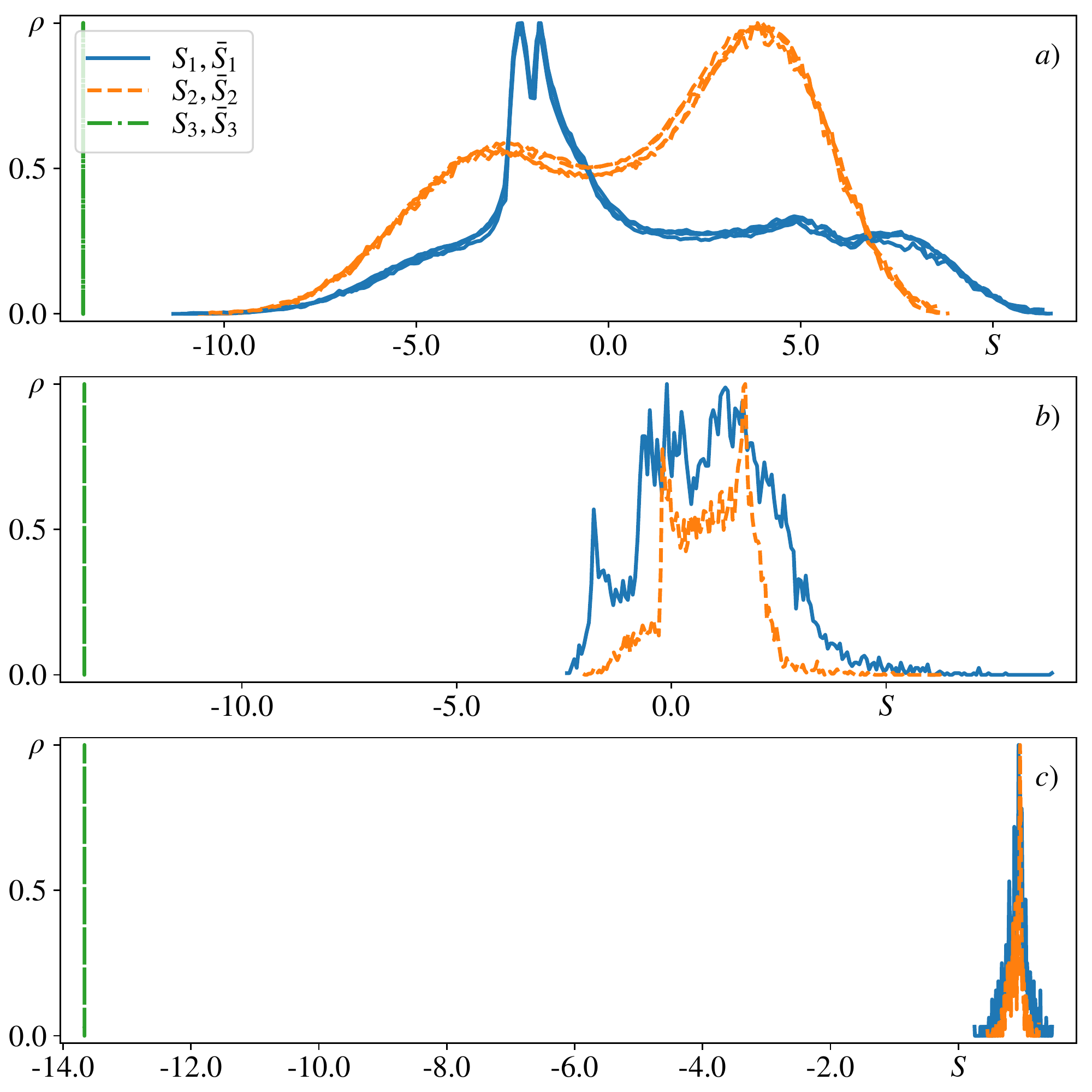}
  \caption{\label{fig:lorenz_acm}Distributions of summed IBLEs and
    FTLEs, see Eq.~\eqref{eq:ible_acm}, for the Lorenz
    system~\eqref{eq:lorenz}: (a) distributions of $S_n$ and
    $\bar{S}_n$ computed with numerical step sizes $\numdt=0.01$ and
    $0.001$ (for $\bar{S}_n$ these $\numdt$ are also used as averaging
    times); (b,c) distributions of $\bar{S}_n$ computed with the
    numerical time step $\numdt=0.01$ and with the averaging times
    $\ftledt=1$ and $10$, respectively. Observe that the $\bar{S}_2$
    becomes strictly positive only in panel (c).}
\end{figure}

Figure~\ref{fig:lorenz_acm}a shows the distributions of $S_n(t)$ and
$\bar{S}_n(t,t+\numdt)$, where $\numdt$ is a numerical discretization
time step. The curves have been computed with $\numdt=0.01$ and
$0.001$ so that four curves are plotted at each $n$. According to
Eq.~\eqref{eq:ftble_integral} $\ftble_i(t,t+\numdt)\approx\ible_i(t)$
if the averaging time $\numdt$ is so small that $\ible_i(t)$ varies
slowly on the integration interval. Thus, the coincidence of the
distributions for $S_n(t)$ and $\bar{S}_n(t,t+\numdt)$ indicates that
the instant exponents $\ible_i(t)$, though computed for a discrete
subset of trajectory points, catch nevertheless all its essential
features. On the other hand, $\ftble_i(t,t+\numdt)$ being averaged
over time step, nevertheless does not ignore essential fine
details. Moreover the coincidence of the distributions for different
discretization steps $\numdt$ indicates that these results are not
affected by numerical approximation errors. Altogether, the
coincidence of the four distributions for each $n$ guarantees that
they are representative, i.e., adequately reveal instant volume
expanding properties of the attractor.

\begin{figure}[tbp]
  \onefig{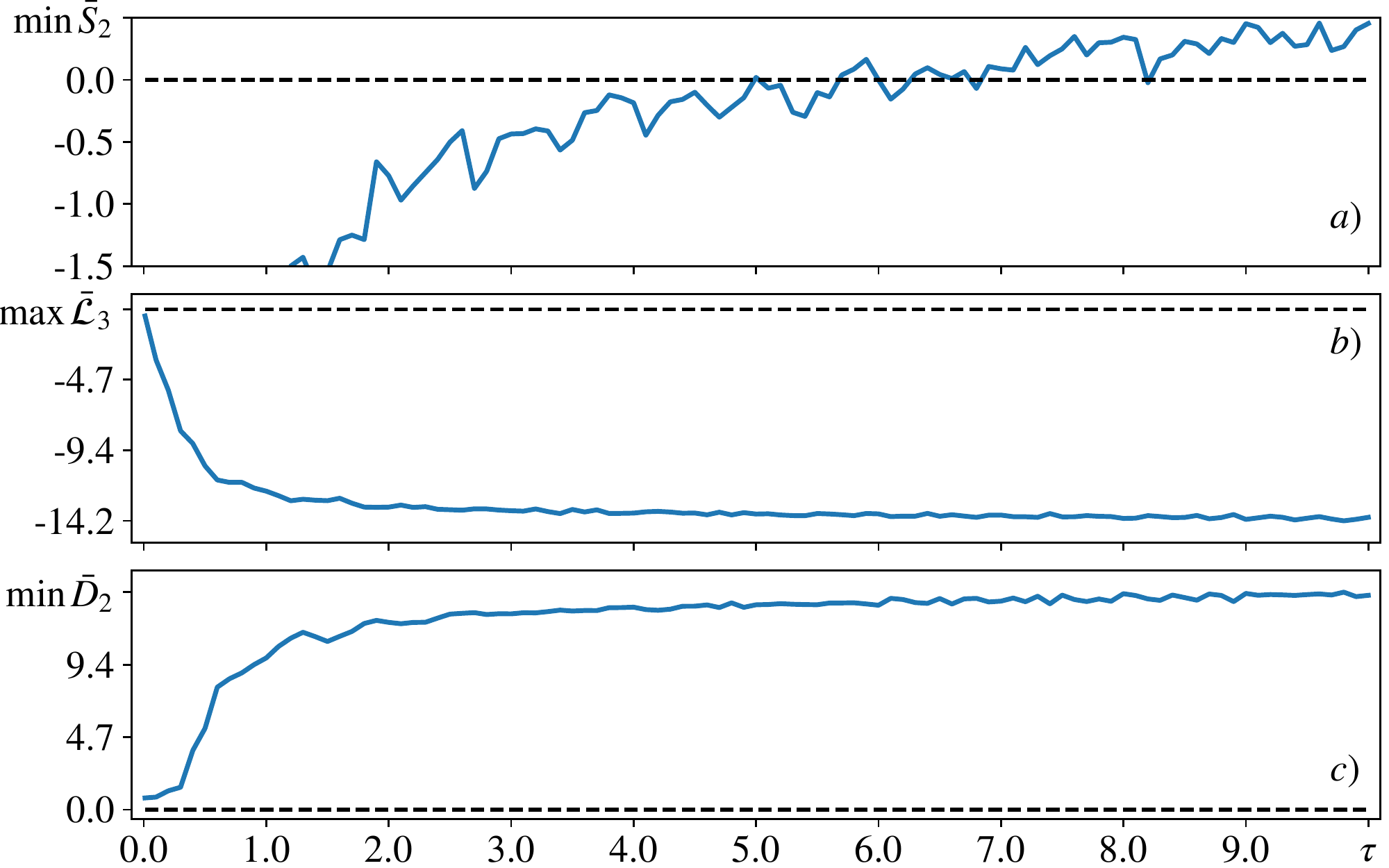}
  \caption{\label{fig:lorenz_minmax}Boundaries of distributions
    vs. averaging time $\ftledt$ for the Lorenz
    system~\eqref{eq:lorenz}: (a) the lower boundary of the
    distribution of $\bar{S}_2(t,t+\ftledt)$, (b) the upper boundary
    of $\ftcle_3(t,t+\ftledt)$, and (c) the lower boundary of the
    distribution of distances the between FTCLEs
    $\bar{D}_2(t,,t+\ftledt)$. Dashed horizontal line shows zero
    level. Observe that $\minval\bar{S}_2$ becomes positive only at
    sufficiently large time scale.}
\end{figure}

The curve $S_2$ in Fig.~\ref{fig:lorenz_acm} is responsible for the
tested property \ref{it:psh_expand}. One can see in
Fig.~\ref{fig:lorenz_acm}a that $S_2$ can be both positive and
negative. It means that the first subspace being expanding on average
due to $\lambda_1+\lambda_2>0$ on infinitesimal time can be both
volume expanding and contracting. Figures.~\ref{fig:lorenz_acm}b and
(c) shows the distributions of sums of FTLEs $\bar{S}_n$ computed for
finite times $\ftledt=1$ and $10$, respectively. Practically we
average every hundred and every thousand of FTLEs, respectively,
computed with the time step $\numdt=0.01$. One can see that only in
panel (c) the distribution of $\bar{S}_2$ becomes strictly
positive. Thus the first subspace becomes expanding only on
sufficiently large time scale. Figure~\ref{fig:lorenz_minmax}a
illustrates it in more detail. It shows behavior of the lower boundary
of the distribution of $\bar{S}_2$ vs. the averaging time
$\ftledt$. The property~\ref{it:psh_expand} is fulfilled when
$\minval \bar{S}_2>0$ at roughly $\ftledt>7$.

In Fig.~\ref{fig:lorenz_acm} the distribution of $S_3$, the sum of all
exponents showing the volume contraction in the whole tangent space,
form the $\delta$ peak. This contraction exponent is known to be equal
to the divergence of the vector field generated by
Eq.~\eqref{eq:lorenz} and is equal to $-(\sigma+b+1)$, i.e., is
constant for each trajectory. For the particular values of parameters,
the divergence is $-41/3\approx -13.67$. Analysis of the data used for
plotting Fig.~\ref{fig:lorenz_acm} shows that $S_3$ as expected is
always constant and equal to this value, so that its distribution
always form the $\delta$ peak.

\begin{figure}[tbp]
  \onefig{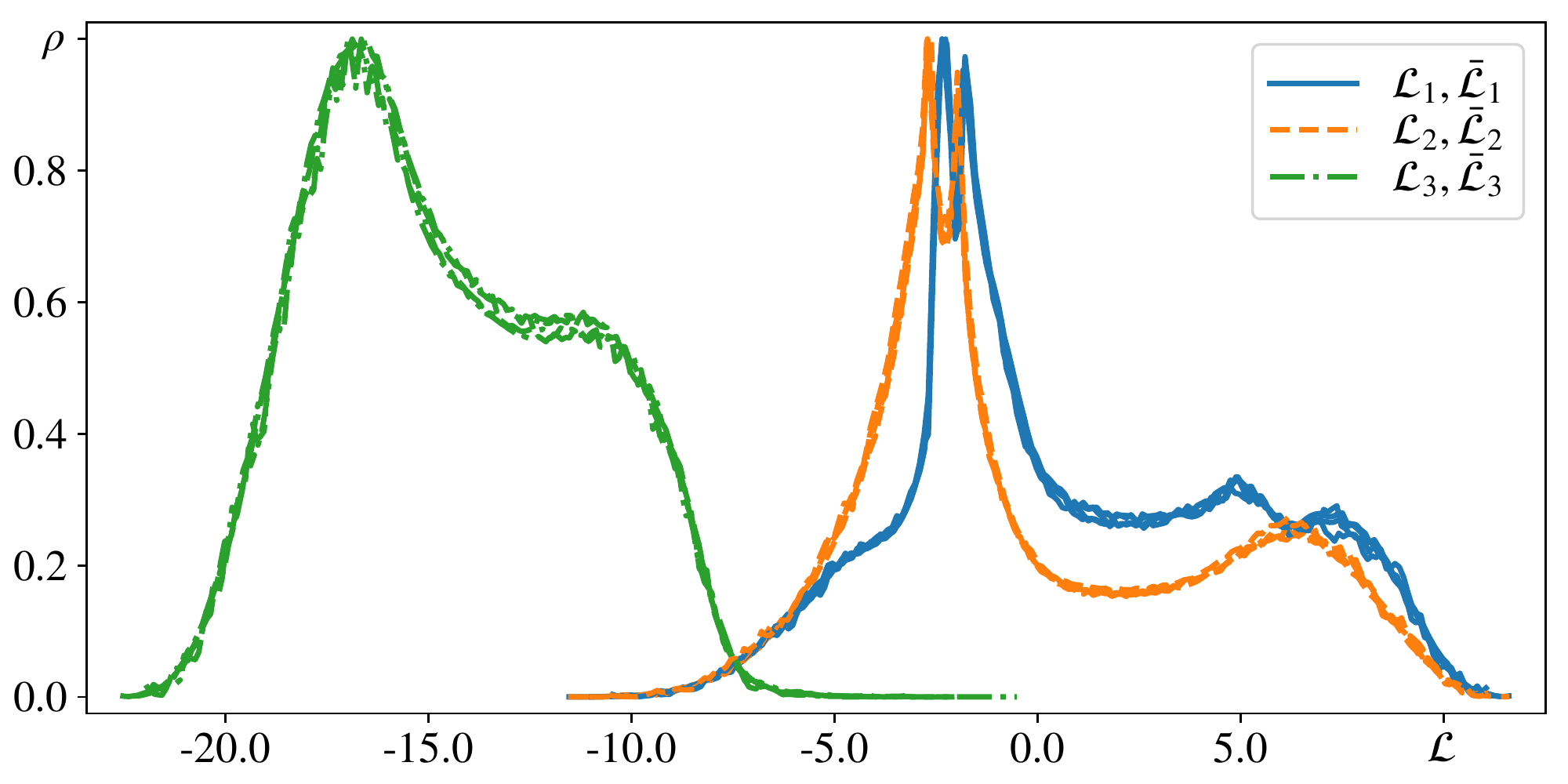}
  \caption{\label{fig:lorenz_icle}Distributions of ICLEs $\icle_i(t)$
    and the corresponding finite time exponents FTCLEs
    $\ftcle_i(t,t+\numdt)$ for the Lorenz system~\eqref{eq:lorenz}
    computed with numerical steps $\numdt=0.01$ and $0.001$. The
    distribution for $\icle_3$ is fully localed on the negative
    semiaxis so that the second subspace is contracting already on
    infinitesimal times.}
\end{figure}

The verification of the property \ref{it:psh_contr}, that the second
subspace is contracting, can be done with the help of ICLE
$\icle_n(t)$. Similar to Fig.~\ref{fig:lorenz_acm}a in
Fig.~\ref{fig:lorenz_icle} we plot the distributions of $\icle_n(t)$
computed with numerical steps $\numdt=0.01$ and $0.001$ and also the
distributions of the corresponding FTCLEs $\ftcle_n(t,t+\numdt)$. The
coincidence of the four curves for each exponent index $n$ guarantees
that the distributions are representative. The contraction in the
second subspace is given by $\icle_3(t)$. Observe that it is always
negative, so that the property~\ref{it:psh_contr} is fulfilled already
on infinitesimal time. Figure~\ref{fig:lorenz_minmax}b illustrates that this
property is fulfilled on finite times. One can see that the upper
boundary of the distribution $\maxval\ftcle_3$ goes lower to negative
area as the averaging time $\ftledt$ grows.

To test if any contraction in the first subspace is exponentially
weaker than the contraction in the second one
(property~\ref{it:psh_dosmmx}), we consider the distribution of
distances between ICLEs $D_n$ and FTCLEs $\bar{D}_n$:
\begin{equation}
  \label{eq:dosmmx}
  \begin{aligned}
    D_n(t)&=\minval\{\icle_i(t),1\leq i\leq
    n\}\\&-\maxval\{\icle_i(t),n+1\leq i\leq N\},\\
    \bar{D}_n(t,t+\ftledt)&=\minval\{\ftcle_i(t,t+\ftledt),1\leq i\leq
    n\}\\&-\maxval\{\ftcle_i(t,t+\ftledt),n+1\leq i\leq N\}.
  \end{aligned}
\end{equation}
This characteristic value is similar to the so called fraction of DOS
violation criterion which implies pairwise comparison of FTCLEs and
counting situations when $\ftcle_i<\ftcle_j$, where $j>i$. Here the
abbreviation DOS stands for dominated Oseledec splitting. This
characteristic value is used in
Refs.~\cite{EffDim,ModesSplit10,EffDim2} to verify the hyperbolic
isolation of tangent modes in spatially distributed systems.

\begin{figure}[tbp]
  \onefig{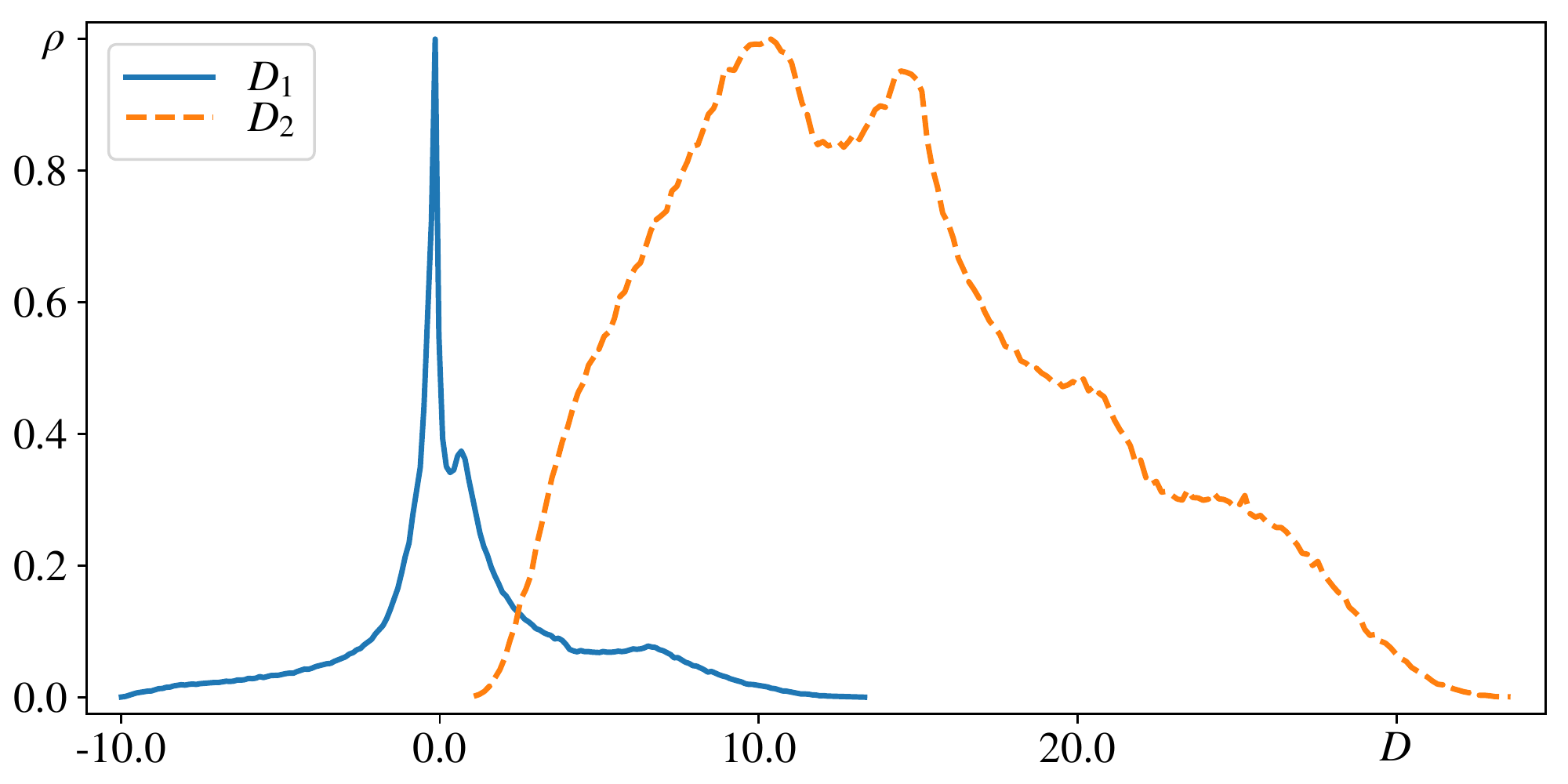}
  \caption{\label{fig:lorenz_dosmmx}Distributions of distances between
    ICLEs $D_n$, Eq.~\eqref{eq:dosmmx}, computed with time step
    $\numdt=0.01$ for the Lorenz system~\eqref{eq:lorenz}. Observe
    that the distribution for $D_2$ falls onto the positive semiaxis.}
\end{figure}

The splitting between the first and the second subspaces is
characterized by $D_2$, the difference between the smallest ICLE in
the first subspace $\minval\{\icle_1(t),\icle_2(t)\}$, and ICLE from
the second subspace $\icle_3(t)$. One can see in
Fig.~\ref{fig:lorenz_dosmmx} that in agreement with the property
\ref{it:psh_dosmmx} $D_2$ is always positive, so that any instant
contraction in the first subspace is always weaker than instant
contractions in the second subspace. Figure~\ref{fig:lorenz_minmax}c
shows that this property is fulfilled on finite time scales. One can
see that the smallest distance $\minval\bar{D}_2$ between the finite
time exponents FTCLEs goes to positive area as averaging time
$\ftledt$ grows.

Fluctuations around zero of $D_1$ in Fig.~\ref{fig:lorenz_dosmmx}
indicate that inside the first subspace the first exponent $\icle_1(t)$
can often be smaller than the second one $\icle_2(t)$. These strong
fluctuations result in the high entanglement of the corresponding
covariant vectors and vanishings of the angle $\theta_1$ in
Fig.~\ref{fig:lorenz_ang}.

Altogether, for pseudohyperbolic Lorenz attractor we observe that the
tangent space is split into two subspaces, two and one-dimensional,
respectively. These subspaces are hyperbolically isolated from each
other (property \ref{it:psh_ang}). The second subspace is strictly
contracting (property \ref{it:psh_contr}) even on infinitesimal
times. If some contraction occurs in the first subspace, it is weaker
than the contraction in the second subspace
(property~\ref{it:psh_dosmmx}), and this property is also fulfilled
already on infinitesimal time.  But as for the
property~\ref{it:psh_expand}, that the first subspace always expands
volumes, it is fulfilled only when the volume expansion is considered
on sufficiently large time scales.

\subsection{R\"ossler System}
As a counter example where the pseudohyperbolicity is absent we
consider the well known R\"ossler
system~\cite{Roess76,SchusJust05,KuzDynChaos06}
\begin{equation}\label{eq:roessler}
  \begin{aligned}
    \dot x &= -y-z,\\
    \dot y &= x+ay,\\
    \dot z &= b+z(x-c),
  \end{aligned}
\end{equation}
with parameters $a=0.2$, $b=0.2$, $c=5.7$.

Lyapunov exponents computed with time step $\numdt=0.0001$ until the
maximal absolute error $10^{-5}$ is reached and averaged over ten
trajectories are $\lambda_1=0.072$,
$\lambda_2\approx 1\times 10^{-6}$, and $\lambda_3=-5.394$. The second
one must be put to zero as being responsible for perturbation along
the trajectory.

\begin{figure}[tbp]
  \onefig{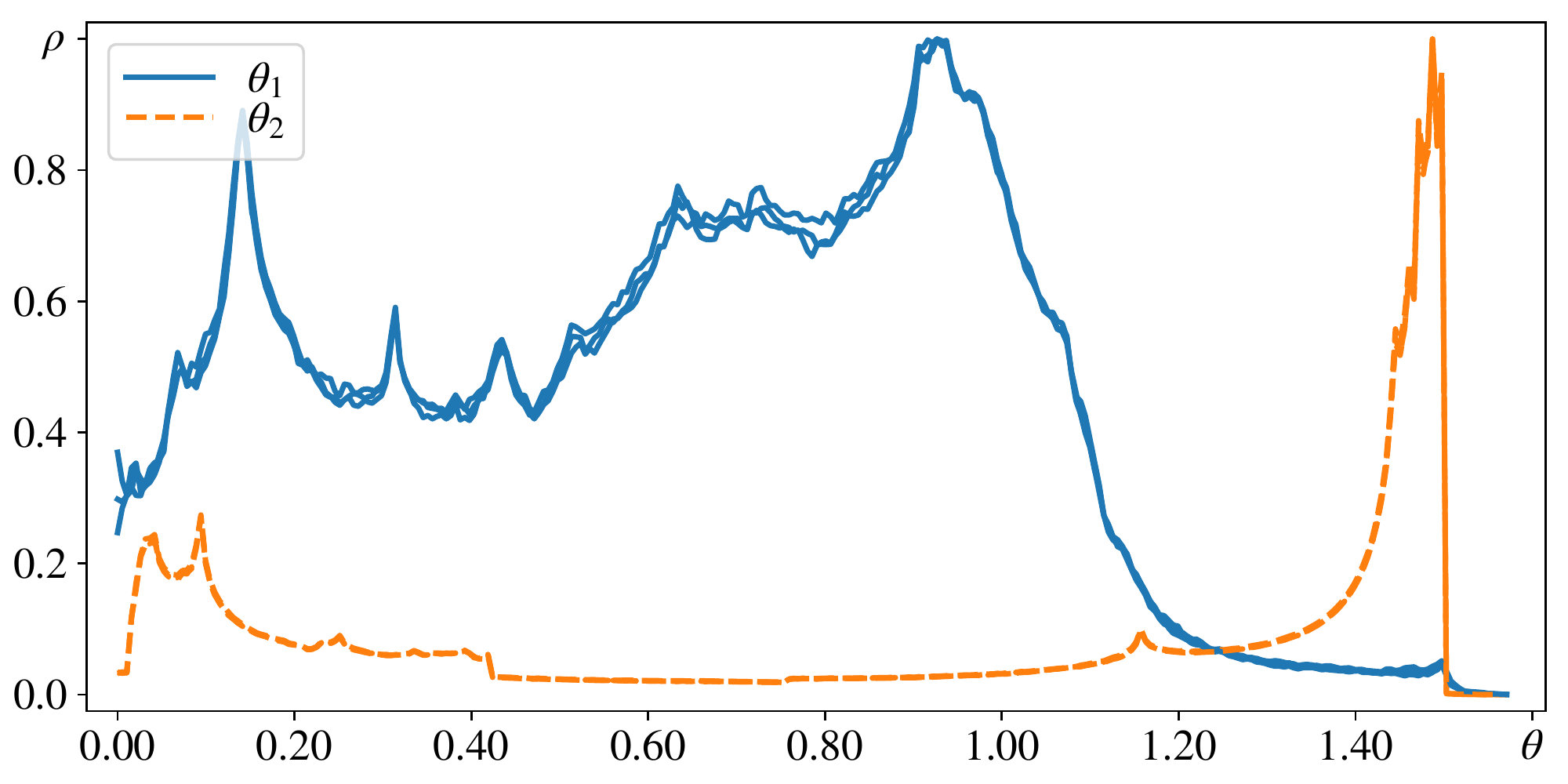}
  \caption{\label{fig:roessler_ang}Distributions of angles between
    tangent subspaces for the R\"ossler
    system~\eqref{eq:roessler}. Each curve is computed three times
    with different numerical steps $\numdt=0.01$, $0.001$ and
    $0.0001$. Both angles often vanish so that there are no
    hyperbolically isolated tangent subspaces.}
\end{figure}

Though the necessary condition for pseudohyperbolicity
$\lambda_1+\lambda_2>0$ is fulfilled, this is not a pseudohyperbolic
attractor since the first two-dimensional tangent subspace is not
hyperbolically isolated from the second one-dimensional subspace: as
shows Fig.~\ref{fig:roessler_ang}, the distribution for the
corresponding angle $\theta_2$ is not separated from zero. The angle
$\theta_1$ can also vanish, so that all tangent subspaces of the
R\"ossler system are highly entangled and no splitting into
hyperbolically isolated subspaces exists.

\begin{figure}[tbp]
  \onefig{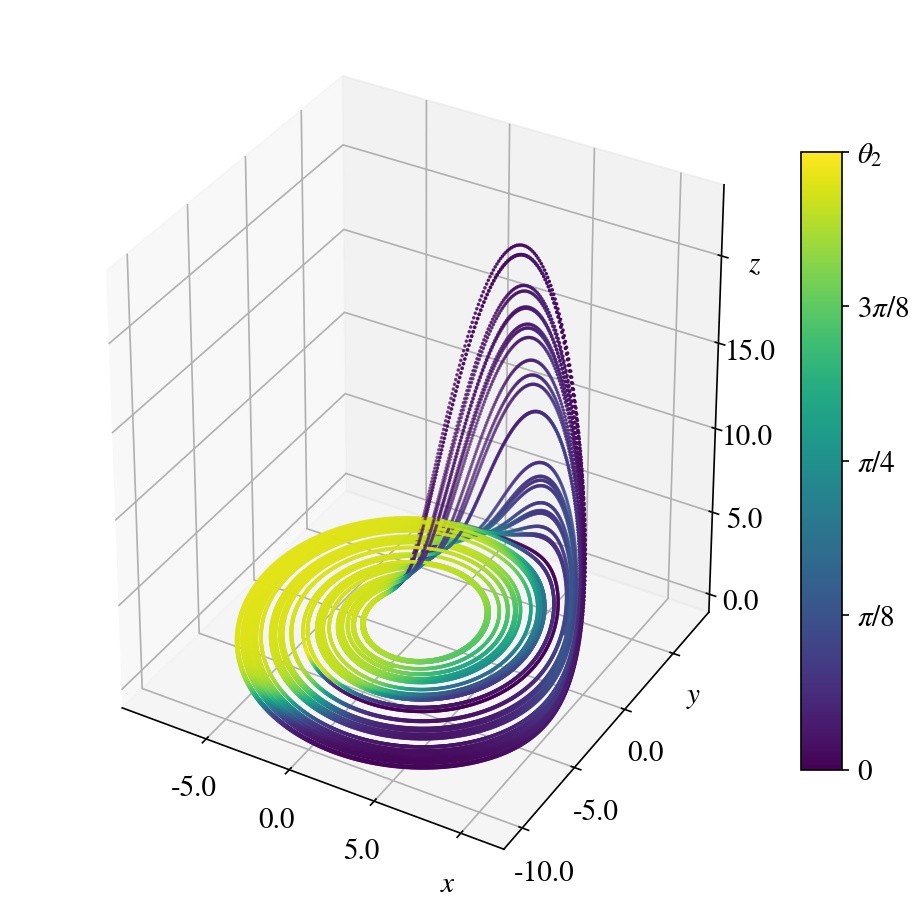}
  \caption{\label{fig:roessler_attr3d}Phase portrait of the R\"ossler
    system~\eqref{eq:roessler}. Points are painted according to the
    values of $\theta_2$. Observe large number of points with
    vanishing angles.}
\end{figure}

Figure~\ref{fig:roessler_attr3d} shows the phase portrait of the
R\"ossler system painted according to values of the angle
$\theta_2$. One can see that the tangencies indicated by zeros of
$\theta_2$ (dark areas) occupies a half of the circle-like horizontal
band laying parallel to $xy$-plane, and also $\theta_2$ vanishes along
loops going up along $z$-axis.

\begin{figure}[tbp]
  \onefig{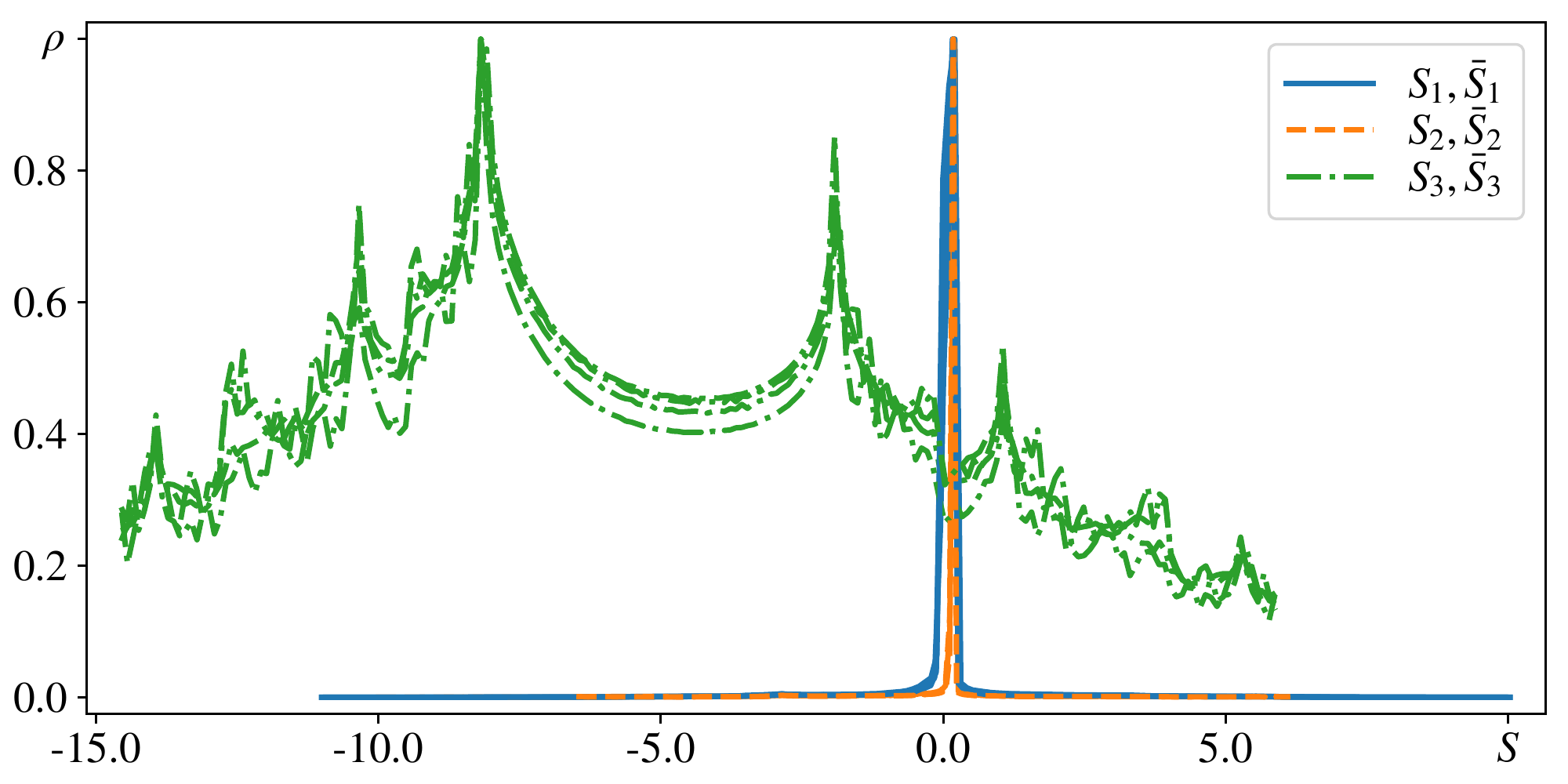}
  \caption{\label{fig:roessler_acm}Distributions of summed IBLEs and
    FTLEs for the R\"ossler system~\eqref{eq:roessler}. For each $n$
    the distributions of $S_n$ and $\bar{S}_n$ are computed with
    numerical step sizes $\numdt=0.01$ and $0.001$. Observe that for
    all $n$ the distributions have both positive and negative tails.}
\end{figure}

We also have tested related properties of the R\"ossler
system~\eqref{eq:roessler}. Figure~\ref{fig:roessler_acm} shows the
distributions of summed IBLEs $S_n(t)$ and FTLEs
$\bar{S}_n(t,t+\numdt)$ indicating the volume expansion
(property~\ref{it:psh_expand}). As above for the Lorenz system, for
each $n$ the distributions are computed with numerical steps
$\numdt=0.01$ and $0.001$. The corresponding curves are barely
distinguishable thus confirming that they are appropriate for
representation of instant expansion and contraction properties.  We
can see that the curves for each $n$ have tails both in positive and
negative semiaxes. They are very low for $S_1$ and $S_2$, while $S_3$,
that is responsible for the contraction in the whole tangent space,
oscillates hard. Consequently, non of the tangent subspaces is
strongly contracting or expanding on infinitesimal times.

\begin{figure}[tbp]
  \onefig{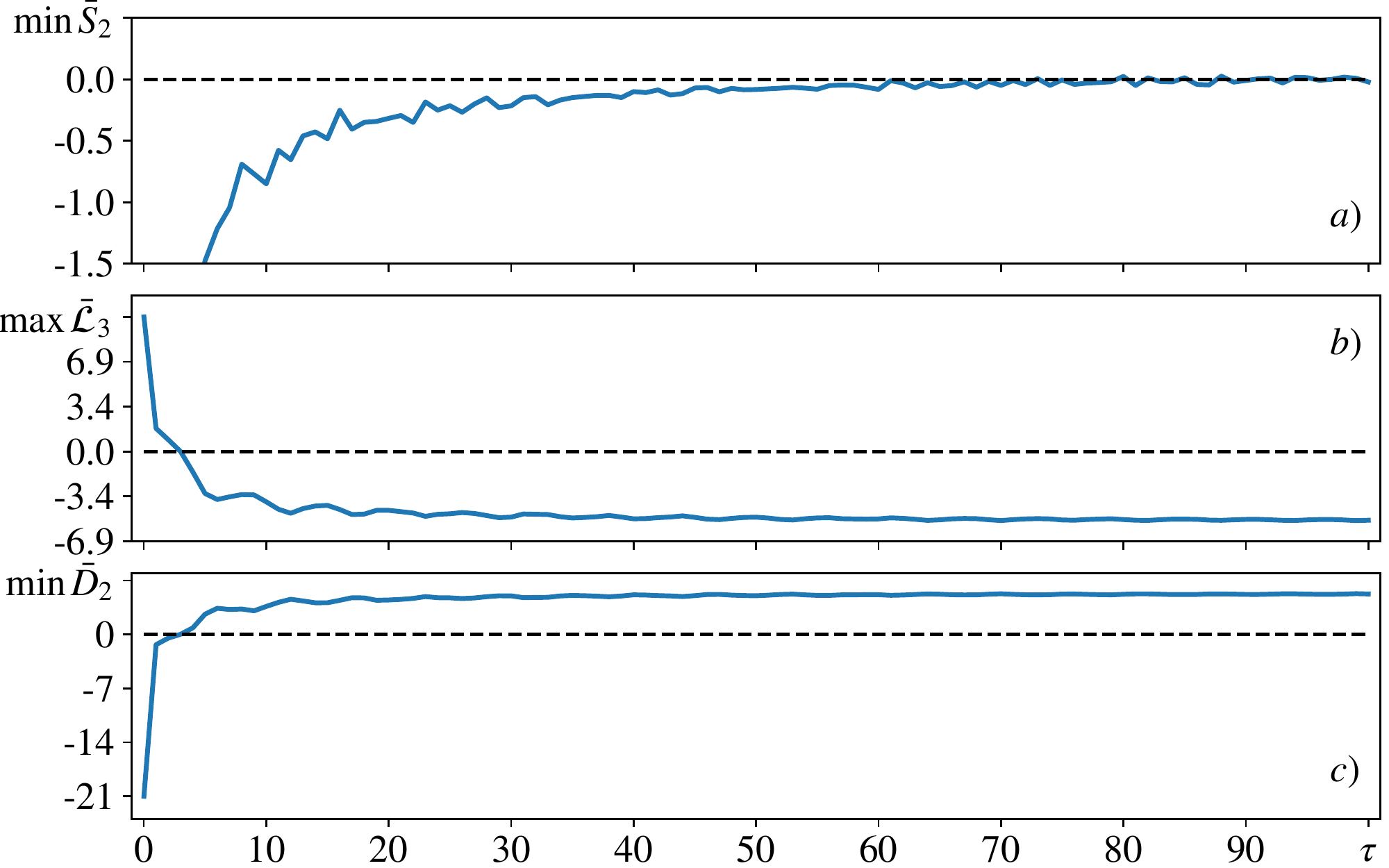}
  \caption{\label{fig:roessler_minmax}As in
    Fig.~\ref{fig:lorenz_minmax} for the R\"ossler
    system~\eqref{eq:roessler}.}
\end{figure}

Figure~\ref{fig:roessler_minmax}a shows behavior of the lower
boundary of the distribution of $\bar{S}_2$ indicating the fulfillment
of the property~\ref{it:psh_expand}. One can see that
$\minval\bar{S}_2$ is negative and the property concerning the volume
expansion remains violated even at sufficiently large time scales.

\begin{figure}[tbp]
  \onefig{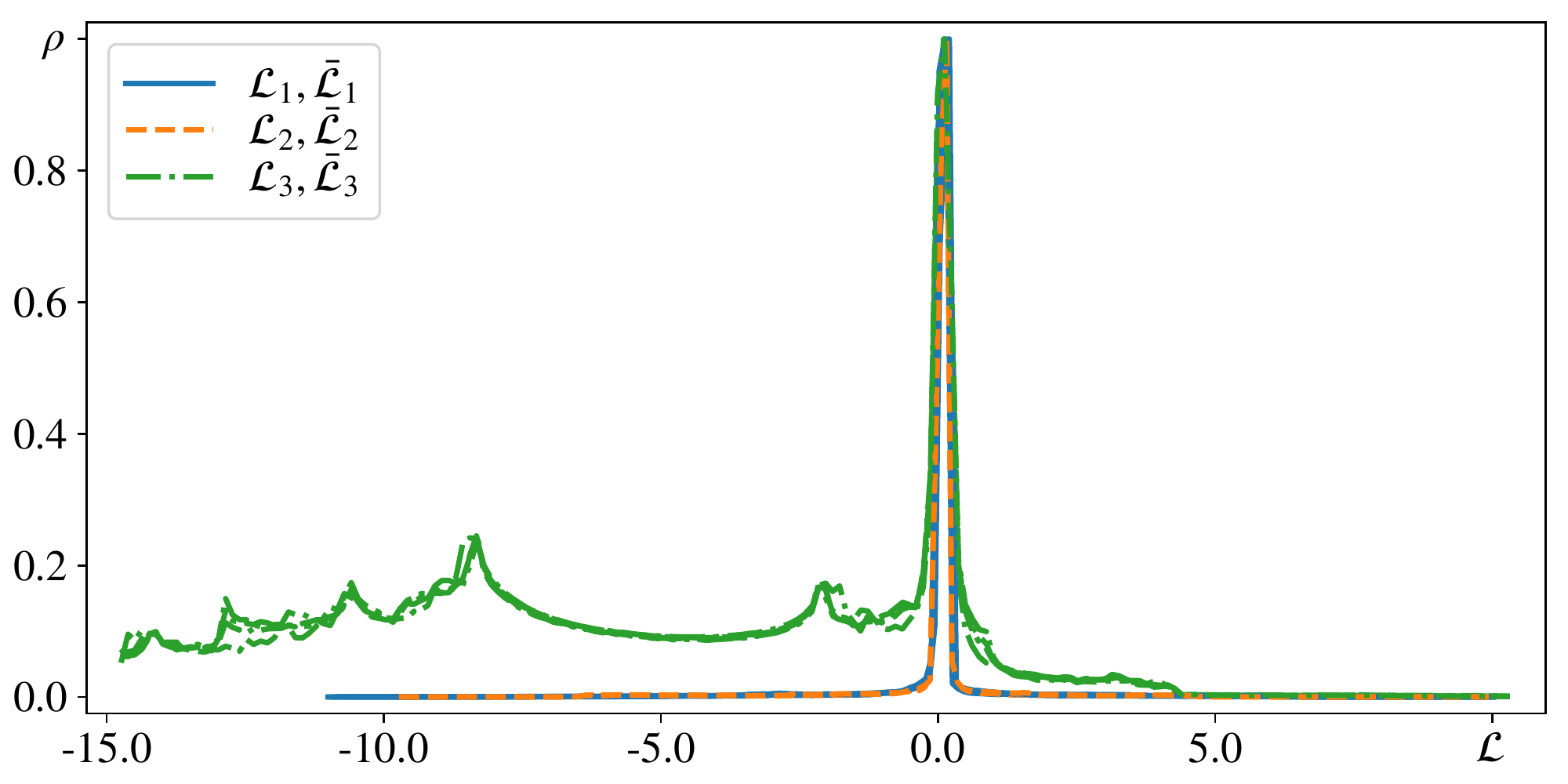}
  \caption{\label{fig:roessler_icle}Distributions of ICLEs
    $\icle_i(t)$ and FTCLEs $\ftcle_i(t,t+\numdt)$ for the R\"ossler
    system~\eqref{eq:roessler}. For each index $n$ the distributions
    are computed using numerical steps $\numdt=0.01$ and $0.001$.
    Observe location of all distributions both on positive and
    negative semiaxes.}
\end{figure}

Figure~\ref{fig:roessler_icle} shows the distributions of ICLEs
$\icle_n(t)$ and the related finite time exponents FTCLEs
$\ftcle_n(t,t+\numdt)$ to verify the contraction in the second
subspace, property~\ref{it:psh_contr}. Again we observe that the
exponents fluctuate around zero so that any covariant direction in the
tangent space on infinitesimal time can be either expanding or
contracting. Figure~\ref{fig:roessler_minmax}b, nevertheless shows
that $\maxval\ftcle_3$ becomes negative at approximately $\ftledt>5$
so that the property~\ref{it:psh_contr} gets fulfilled.

\begin{figure}[tbp]
  \onefig{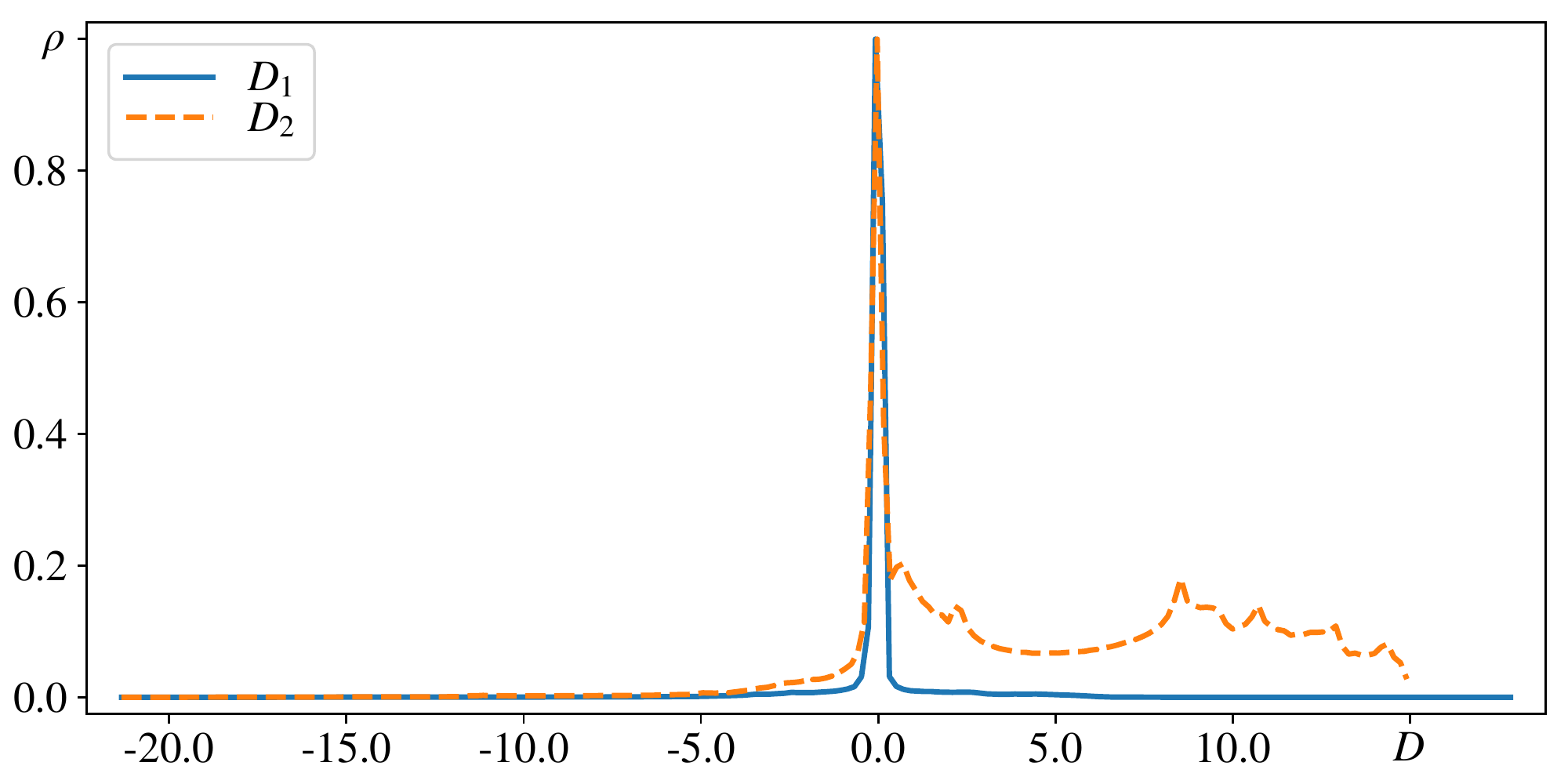}
  \caption{\label{fig:roessler_dosmmx}Distributions of distances
    between ICLEs for the R\"ossler system~\eqref{eq:roessler} solved
    numerically with step size $\numdt=0.01$. Observe that both $D_1$
    and $D_2$ can be both positive and negative.}
\end{figure}

The distributions of distances between ICLEs~\eqref{eq:dosmmx} are
shown in Fig.~\ref{fig:roessler_dosmmx}. The positive and negative
tails of $D_1$ and $D_2$ indicate that on infinitesimal times the
exponents are highly entangled and their order is not preserved. But
as follows from Fig.~\ref{fig:roessler_minmax}c $\minval D_2$
becomes positive at finite time $\ftledt>5$, so that the contraction
in the second subspace becomes strictly stronger than in the first
one, and the property \ref{it:psh_dosmmx} turns fulfilled.

So, the non-pseudohyperbolic nature of the R\"ossler
system~\eqref{eq:roessler} is confirmed due to vanishings of angles
between tangent subspaces. The strict volume expansion within the
first subspace is not observed even at sufficiently large time
scales. The second subspace is not strictly contracting on
infinitesimal time but acquires this property at finite time
scales. The same is the case for the second subspace that turns to be
strictly contracting on finite time scales.

\subsection{Generalized Lorenz System}

Now we will analyze a generalization of the Lorenz system proposed in
Ref.~\cite{ShSh2001}, see problem C.7.No.86, as a possible candidate
for a system with a wild spiral attractor. Also this system as well as
other examples of spiral chaos are considered in
Refs.~\cite{WildSpiral,GonKazTurNew, BorKazSat2016}.
\begin{equation}\label{eq:wildspir}
  \begin{aligned}
    \dot x &= \sigma(y-x),\\
    \dot y &= x(r-z)  - y,\\
    \dot z &= xy - bz + \mu w,\\
    \dot w &= -bw -\mu z,
  \end{aligned}
\end{equation}
where parameters are $r=25$, $\sigma=10$, $b=8/3$, and $\mu=7$.

\begin{figure}[tbp]
  \onefig{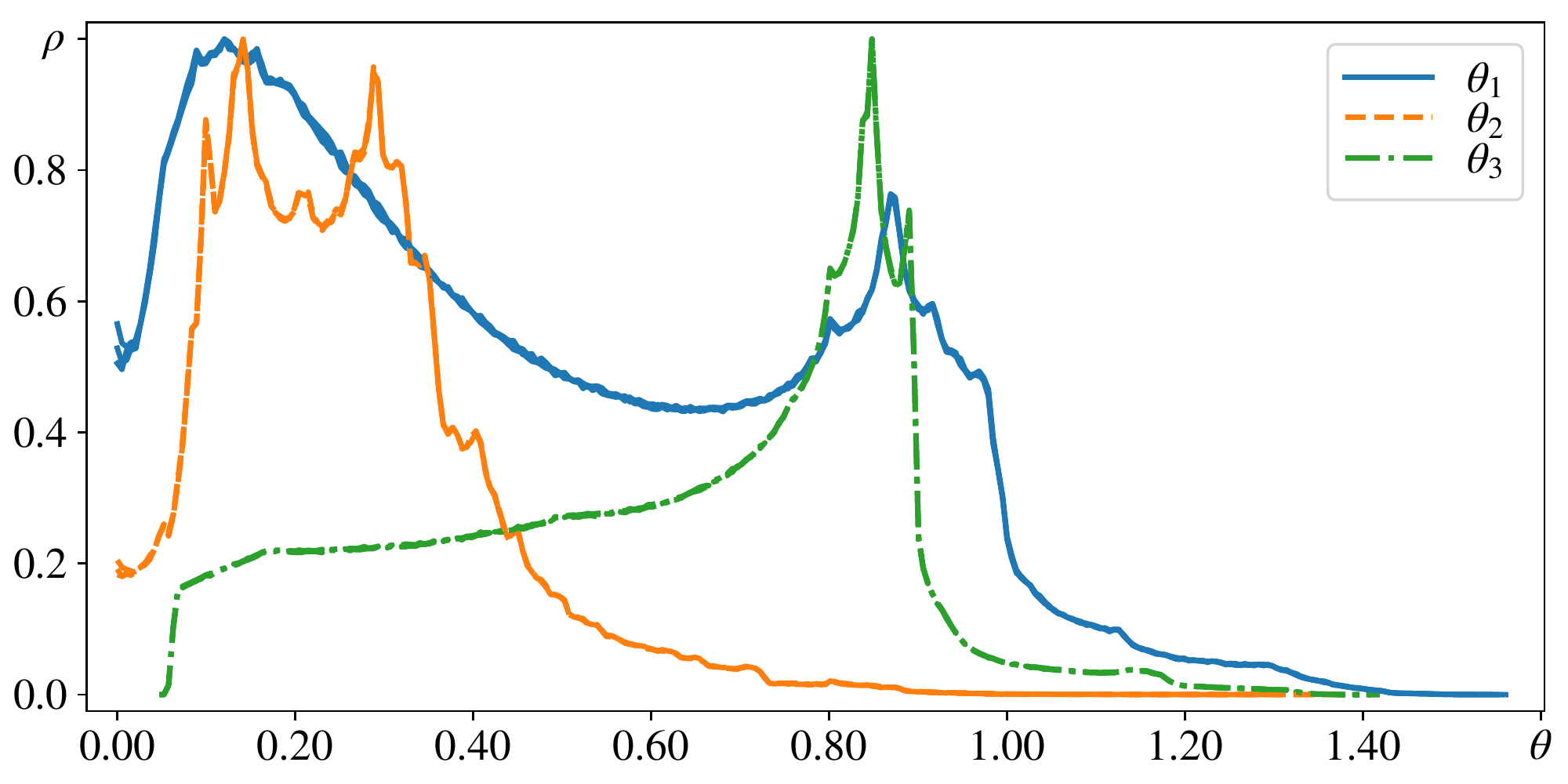}
  \caption{\label{fig:wildspir_ang}Distributions of angles between
    tangent subspaces for the system~\eqref{eq:wildspir}. Each curve
    is computed three times with different numerical steps
    $\numdt=0.01$, $0.001$ and $0.0001$. The pseudohyperbolicity is
    indicated by the distribution of $\theta_3$ that is well detached
    from the origin.}
\end{figure}

Theoretical evaluations suggest that this system is
pseudohyperbolic. Lyapunov exponents computed similarly as for two
previous systems are $\lambda_1=2.193$, $\lambda_2=0$,
$\lambda_3=-1.959$, and $\lambda_4=-16.567$. Since
$\lambda_1+\lambda_2+\lambda_3>0$ and $\lambda_4<0$, the tangent space
splitting responsible for the pseudohyperbolicity, see
property~\ref{it:psh_ang}, is expected to occur between
three-dimensional volume expanding first subspace and one-dimensional
contracting second subspace. Figure~\ref{fig:wildspir_ang} shows the
distributions of angles between the tangent subspaces. The splitting
of interest is characterized by the angle $\theta_3$. Clear separation
of its distribution from the origin confirms that the first and the
second subspaces are hyperbolically isolated so that the
system~\eqref{eq:wildspir} is indeed pseudohyperbolic. Notice also
high frequency of vanishing of $\theta_1$ and $\theta_2$ indicating
that within the first subspace the trajectory manifolds spanned by the
corresponding first three CLVs are highly entangled.

\begin{figure}[tbp]
  \onefig{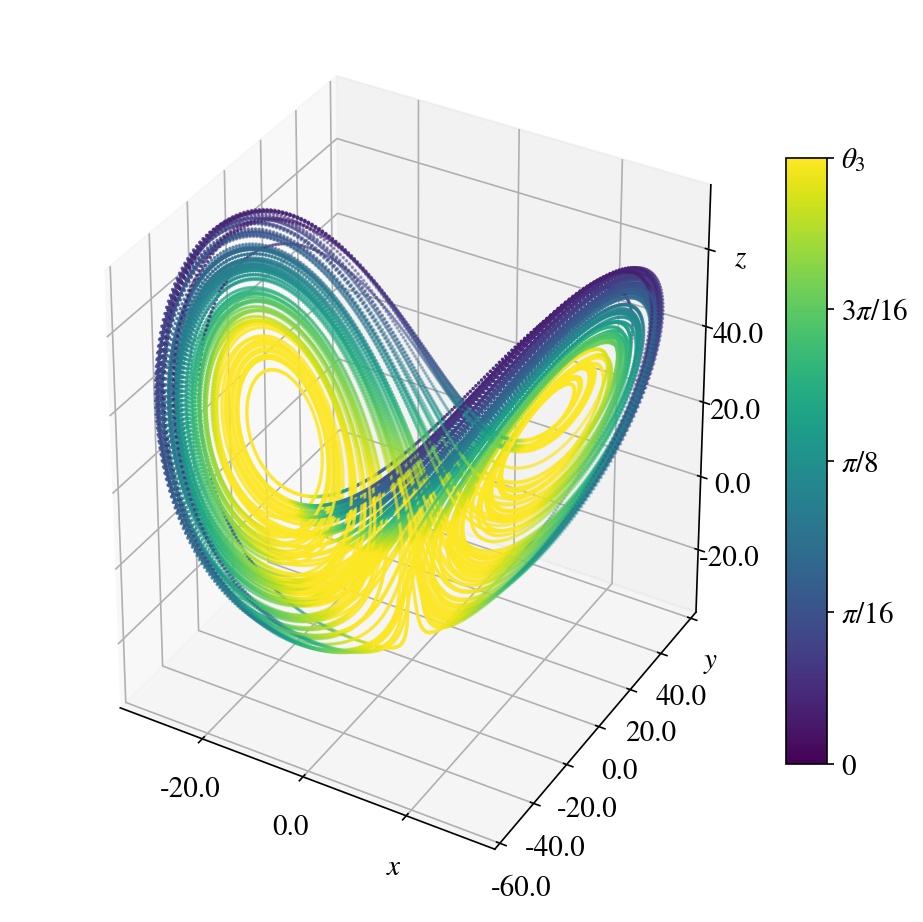}
  \caption{\label{fig:wildspir_attr3d}Attractor for the
    system~\eqref{eq:wildspir}. Colors correspond to values of
    $\theta_3$. Observe similarity with the Lorenz attractor in
    Fig.~\ref{fig:lorenz_attr3d}.}
\end{figure}

Figure~\ref{fig:wildspir_attr3d} shows how values of $\theta_3$ are
located on the attractor of the system~\eqref{eq:wildspir}. It
represents three-dimensional projection of the attractor whose points
are painted according to values of $\theta_3$. Observe that both the
projection itself and the distribution of angles on it is similar to
the Lorenz attractor: it contains two circular bands where small angles
located on outer edges, cf. Fig.~\ref{fig:lorenz_attr3d}.

\begin{figure}[tbp]
  \onefig{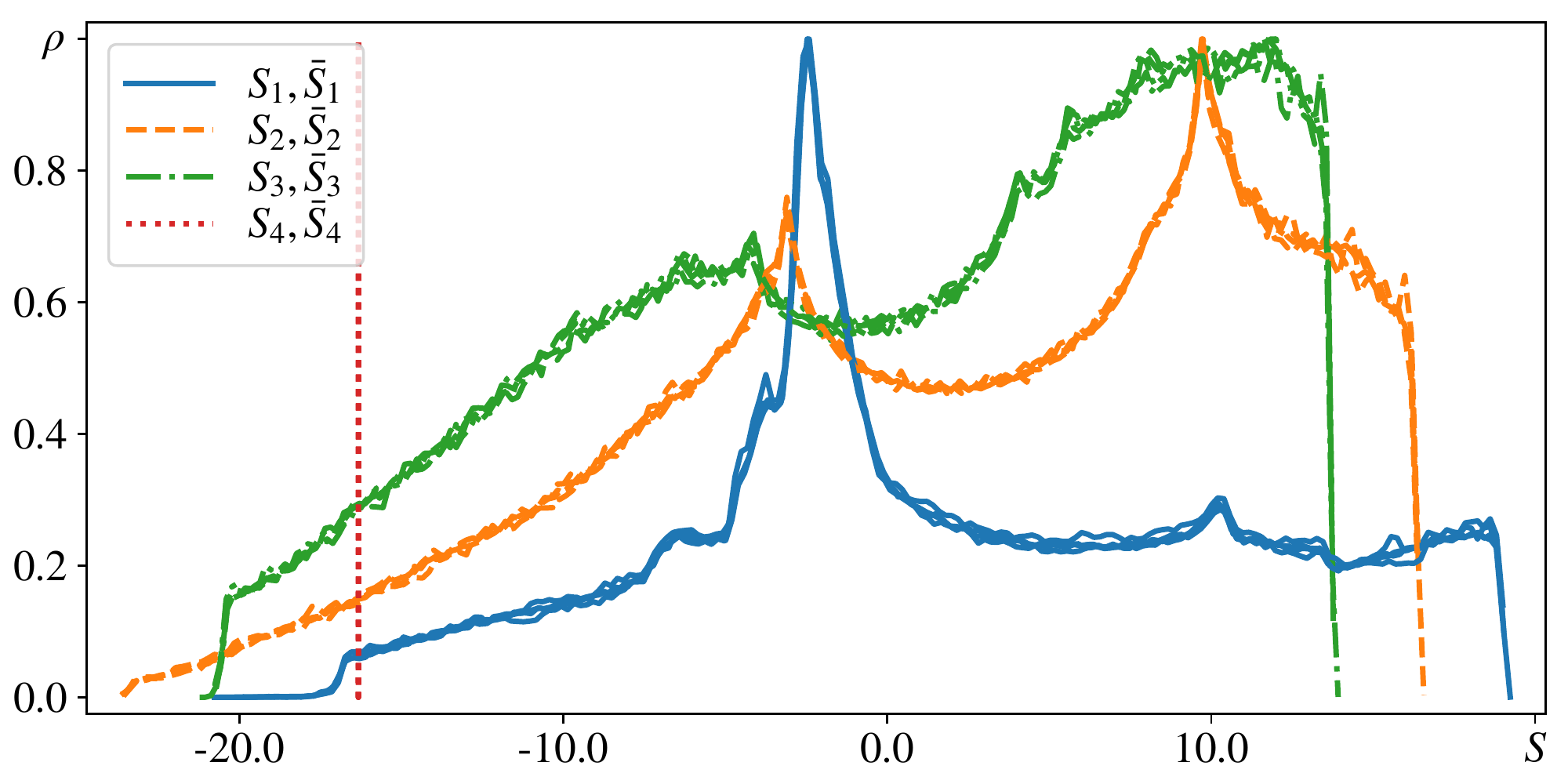}
  \caption{\label{fig:wildspir_acm}Distributions of summed IBLE $S_n$
    and FTLEs $\bar{S}_n$ for the system~\eqref{eq:wildspir} computed
    with numerical steps $\numdt=0.01$ and $0.001$. Observe that $S_3$
    can have both positive and negative signs.}
\end{figure}

Figure~\ref{fig:wildspir_acm} provides verification of the volume
expansion, property~\ref{it:psh_expand}, demonstrating the
distributions of summed exponents $S_n$ and $\bar{S}_n$. As for the
previous systems each curve is computed four times: for IBLEs and
FTLEs with numerical steps $\numdt=0.01$ and $0.001$. Almost perfect
coincidence of the different versions of the curves confirms that they
are representative for characterizing the properties of the attractor
on infinitesimal times.

\begin{figure}[tbp]
  \onefig{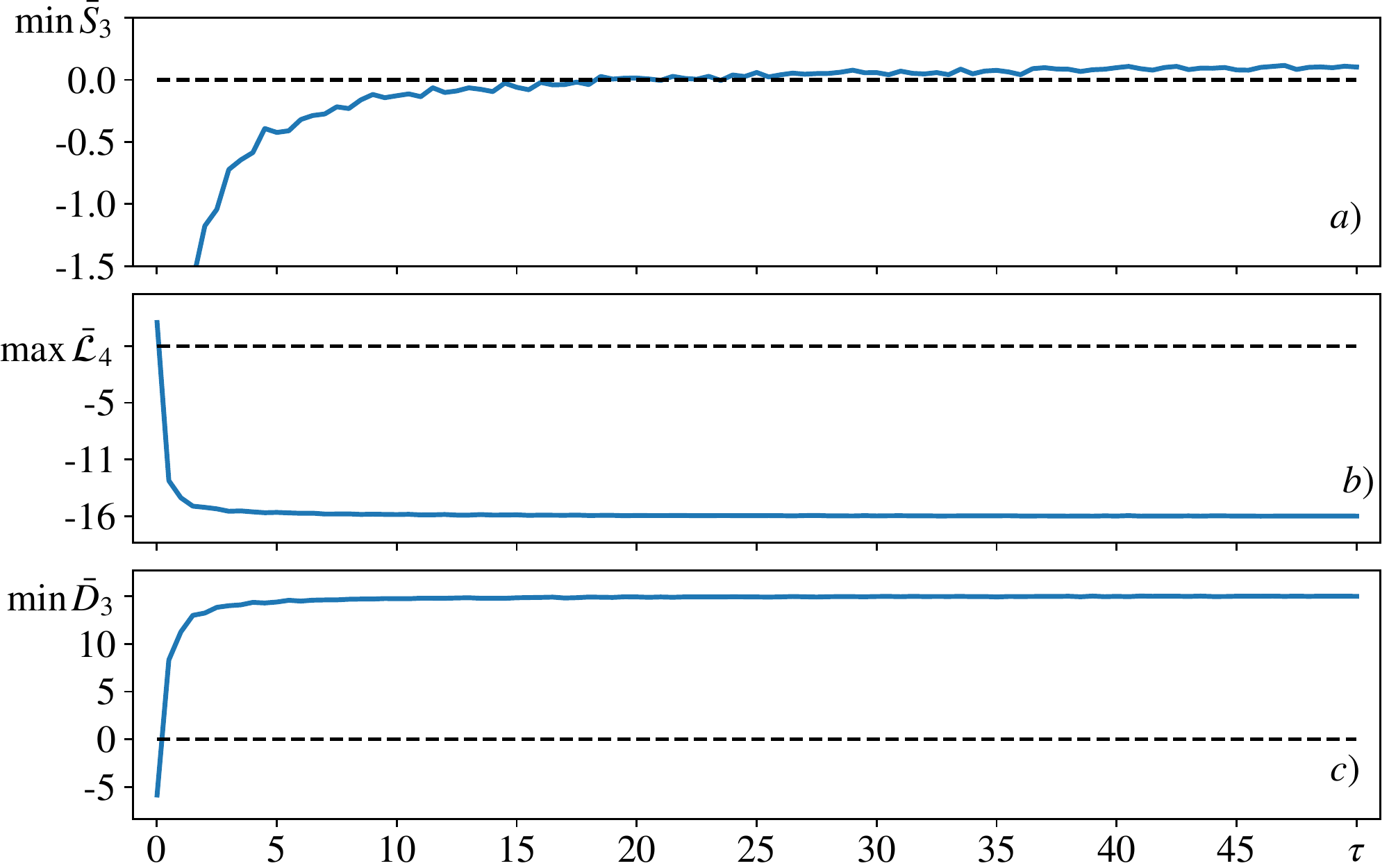}
  \caption{\label{fig:wildspir_minmax}As in
    Fig.~\ref{fig:lorenz_minmax} for the system~\eqref{eq:wildspir}.}
\end{figure}

The volume expansion within the first subspaces is shown by the
distribution for $S_3$. One can see that it hardly oscillates being
with almost equal probabilities both positive and negative. It means
that the property~\ref{it:psh_expand} does not hold on infinitesimal
time. To check when this property becomes fulfilled in
Fig.~\ref{fig:wildspir_minmax} we have plotted the lower boundary of
the distribution $\minval\bar{S}_3$ vs. averaging time $\ftledt$. One
can see that $\minval\bar{S}_3>0$ at approximately $\ftledt>7$. It
means that the first tangent subspace of the
system~\eqref{eq:wildspir} becomes volume expanding at sufficiently
large time scales.

In Fig.~\ref{fig:wildspir_acm} one can see that the distribution for
$S_4$ similarly to the Lorenz system form the $\delta$ peak,
cf. distribution for $S_3$ in Fig.~\ref{fig:lorenz_acm}a. One can
check that this is due to the constant divergence
$\operatorname{div}\fbas=-(\sigma+2b+1)$ that for the given parameter
values is equal to $-16.3$.

\begin{figure}[tbp]
  \onefig{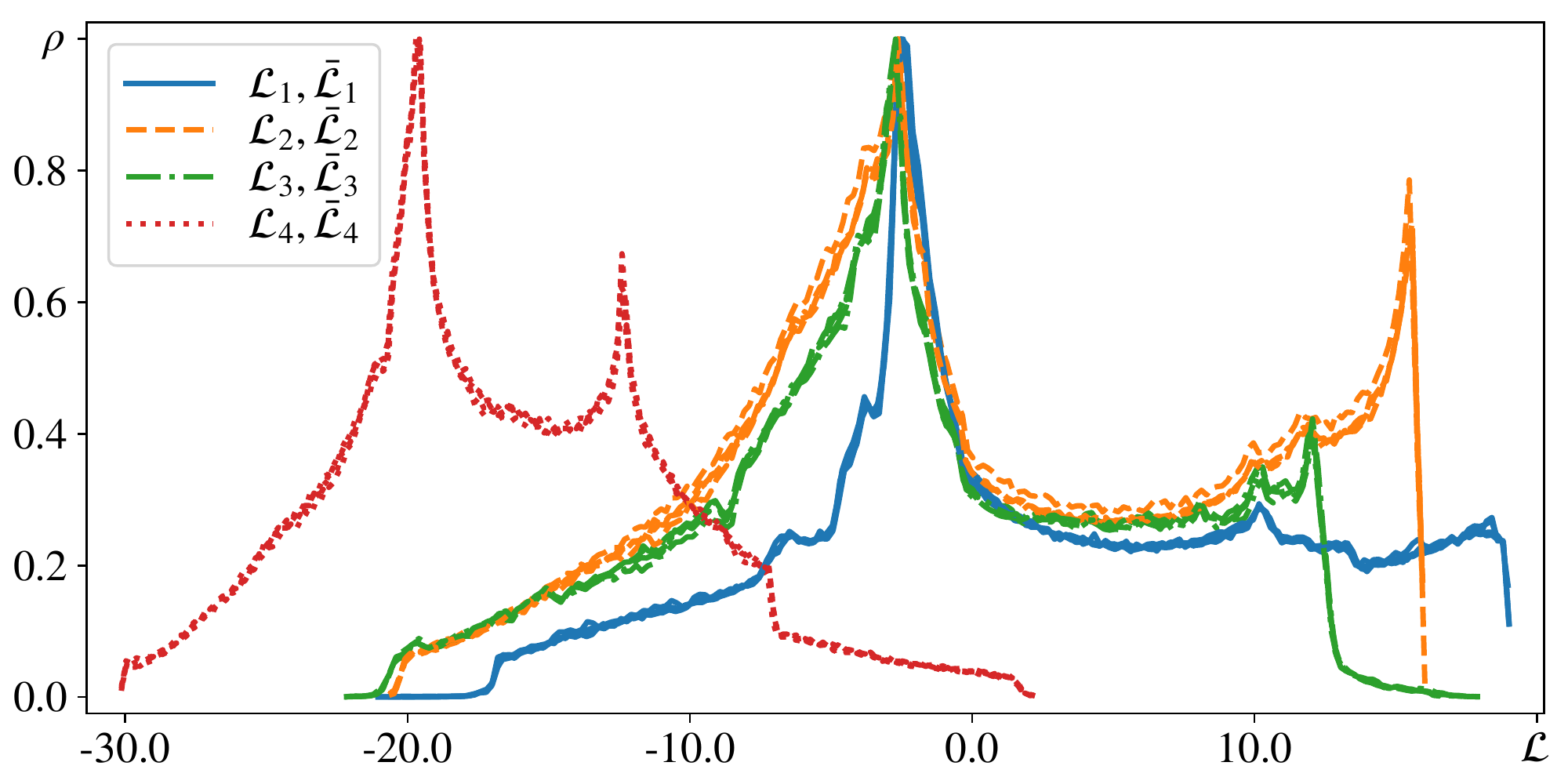}
  \caption{\label{fig:wildspir_icle}(a) Distributions of ICLEs and
    FTCLEs computed with time steps $\numdt=0.01$ and $0.001$ for the
    system~\eqref{eq:wildspir}. Observe that $\icle_4$ can be both
    positive and negative.}
\end{figure}

Property~\ref{it:psh_contr} concerning the strong contraction in the
second subspace is tested in Fig.~\ref{fig:wildspir_icle}. Again each
distribution is represented with four curves: ICLEs and FTCLEs are
computed with numerical time steps $\numdt=0.01$ and $0.001$. One can
see that $\icle_4(t)$ responsible for contraction in the second
subspace, though rarely, can be positive. Therefore, on infinitesimal
time the property~\ref{it:psh_contr} does not hold. As one can see in
Fig.~\ref{fig:wildspir_minmax}b the upper boundary of the
distribution $\maxval\ftcle_4$ turns negative at approximately
$\ftledt>1$, and the second subspace becomes strongly contracting on
finite time scales.

\begin{figure}[tbp]
  \onefig{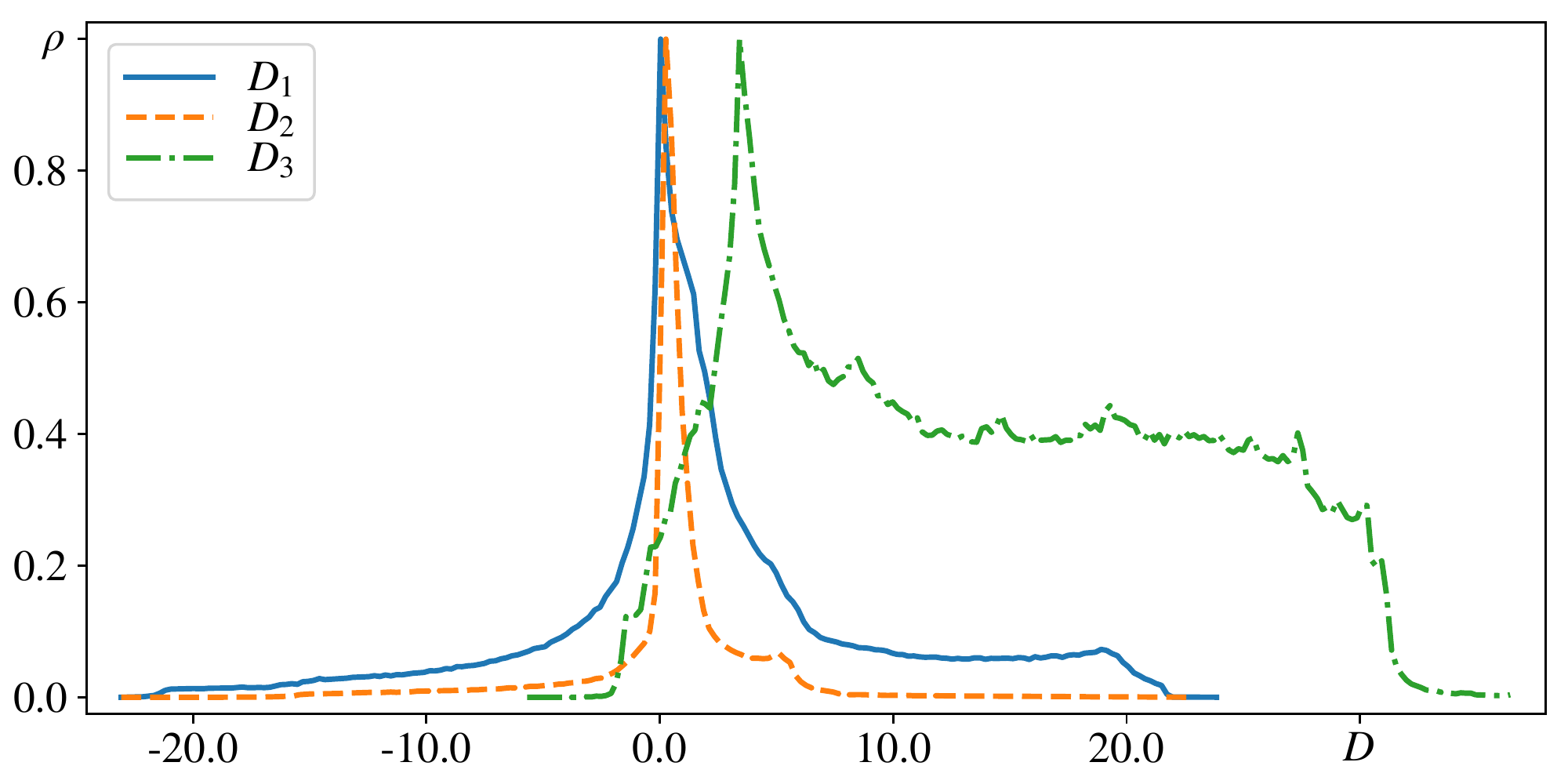}
  \caption{\label{fig:wildspir_dosmmx}Distributions of distances
    between ICLEs for the system~\eqref{eq:wildspir}. Numerical step
    size is $\numdt=0.01$. Observe that $D_3$ changes sign.}
\end{figure}

According to the property~\ref{it:psh_dosmmx}, any contraction in the
first subspace is weaker than contraction in the second subspace.
This is tested using distributions of distances between exponents
$D_n$ in Fig.~\ref{fig:wildspir_dosmmx}. The splitting between the
first and the second tangent subspaces is characterized by $D_3$. One
can see that $D_3$ can rarely be negative. It means that sometimes
instant contraction in the first subspace is stronger than in the
second subspace, and the property \ref{it:psh_dosmmx} does not hold on
infinitesimal times. As follows from Fig.~\ref{fig:wildspir_minmax}c,
the lower boundary $\minval\bar{D}_3$ becomes positive at
approximately $\ftledt>1$, so that the property~\ref{it:psh_dosmmx} is
fulfilled on finite time scales.

Altogether, for generalized the Lorenz system~\eqref{eq:wildspir}, the
pseudohyperbolicity is confirmed due to the absence of tangencies
between the first three-dimensional subspace and the second
one-dimensional subspace, property~\ref{it:psh_ang}. But all other
properties are fulfilled only on finite time scales, and are violated
on infinitesimal times. Volumes from the first subspace can instantly
be contracting and vectors from the second one can sometimes be
expanded. Moreover the instant contraction in the first subspace can
sometimes be stronger than the contraction in the second subspace.

\subsection{Three-Dimensional Generalizations of H\'enon Map}
A series of works have recently been reported where a
pseudohyperbolicity of three-dimensional generalizations of H\'enon
map are discussed~\cite{GonOvs05,GonGonPshyp,PsReview,WildSpiral}. In
this paper we will test the pseudohyperbolicity of the map
\begin{equation}
  \label{eq:henon3d}
  \begin{aligned}
    x_{n+1}&=y_n,\\
    y_{n+1}&=z_n,\\
    z_{n+1}&=Bx_n+Az_n+Cy_n-z_n^2,
  \end{aligned}
\end{equation}
with following parameter sets
\begin{align}
  \label{eq:henon3d_52}
  B&=0.7,\;A=-1.11,\;C=0.77,\\
  \label{eq:henon3d_511}
  B&=0.7,\;A=0,\;C=0.85,\\
  \label{eq:henon3d_512}
  B&=0.7,\;A=0,\;C=0.815.
\end{align}
Parameters~\eqref{eq:henon3d_52} correspond to Eq.~(17) and Fig.~5d
in Ref.~\cite{GonGonPshyp}, and parameters~\eqref{eq:henon3d_511} and
\eqref{eq:henon3d_512} are taken from Ref.~\cite{GonOvs05}, see
Eq.~(1) and Fig.~1 there. 

As reported in Ref.~\cite{WildSpiral}, mathematicians from the
University of Uppsala, Sweden, J. Figueros and W. Tucker using the
interval arithmetic methods have not confirmed the pseudohyperbolicity
of the system~\eqref{eq:henon3d} with parameters
\eqref{eq:henon3d_511} and confirmed it for the
parameters~\eqref{eq:henon3d_512}.

Lyapunov exponents computed with maximal absolute error
$\epsilon=10^{-5}$ and averaged over ten independent trajectories are
the following: for~\eqref{eq:henon3d_52} $\lambda_1=0.013$,
$\lambda_2=0$, $\lambda_3=-0.370$; for~\eqref{eq:henon3d_511}
$\lambda_1=0.020$, $\lambda_2=0$, $\lambda_3=-0.377$; and
for~\eqref{eq:henon3d_512} $\lambda_1=0.008$, $\lambda_2=0$,
$\lambda_3=-0.365$.

\begin{figure}[tbp]
  \onefig{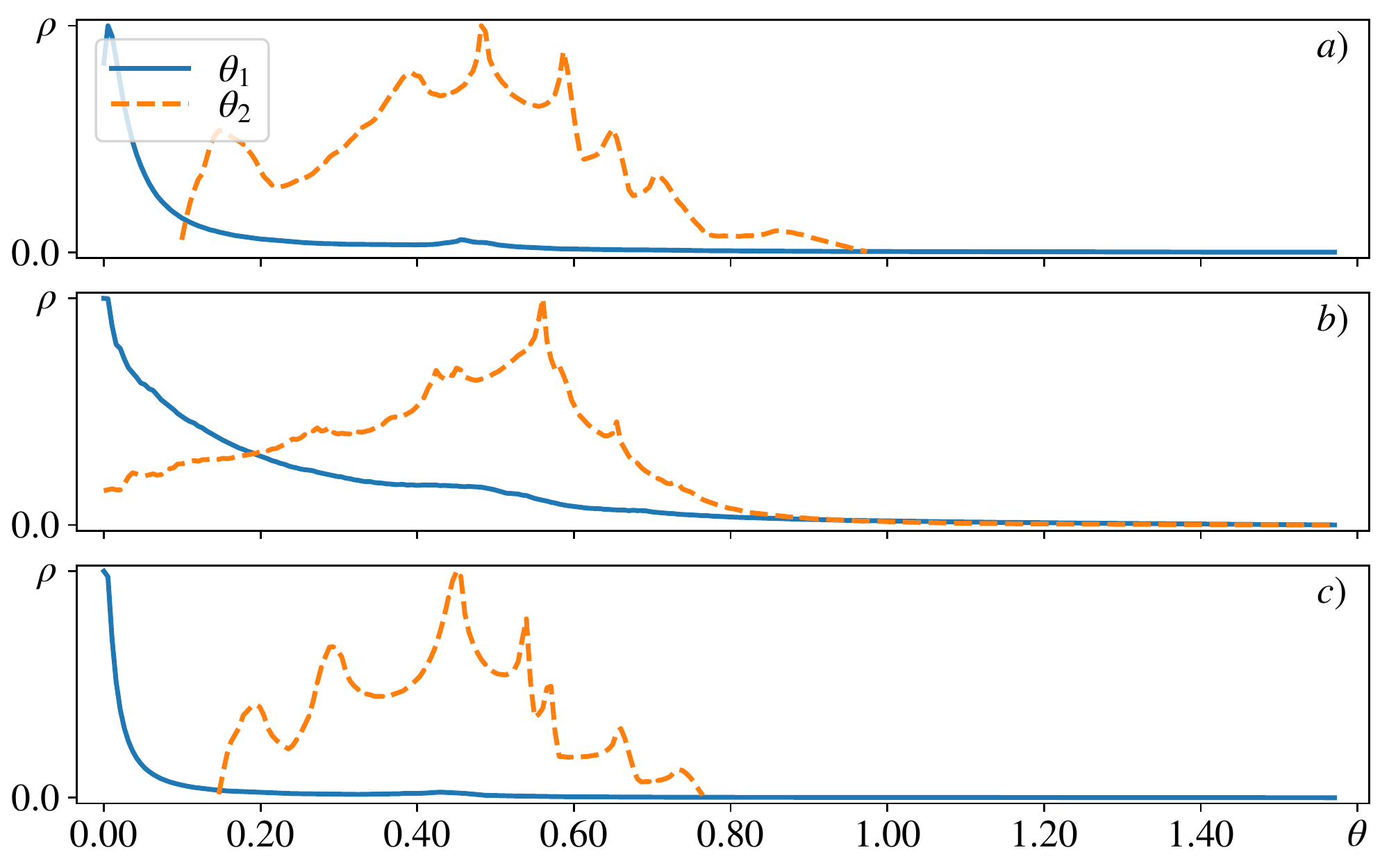}
  \caption{\label{fig:henon3d_all_ang}Distributions of angles between
    tangent subspaces for the system~\eqref{eq:henon3d}. Panels (a),
    (b), and (c) correspond to parameters~\eqref{eq:henon3d_52},
    \eqref{eq:henon3d_511}, and~\eqref{eq:henon3d_512}. Non vanishing
    $\theta_2$ confirms the pseudohyperbolicity in panels (a) and (c),
    while the case represented in panel (b) is not pseudohyperbolic.}
\end{figure}

The presence of the pseudohyperbolicity is tested in
Fig.~\ref{fig:henon3d_all_ang}, where the distributions of angles
between tangent subspaces are shown, property~\ref{it:psh_ang}. Since
for all cases $\lambda_1+\lambda_2>0$ and $\lambda_3<0$, the first
subspace is two-dimensional and the second has one-dimension. It means
that the angle $\theta_2$ indicates the presence or absence of the
pseudohyperbolicity. As one can see in Fig.~\ref{fig:henon3d_all_ang}a
and~\ref{fig:henon3d_all_ang}c, the non vanishing $\theta_2$ indicates
that parameters~\eqref{eq:henon3d_52} and~\eqref{eq:henon3d_512}
corresponds to a pseudohyperbolic attractor, i.e., the
property~\ref{it:psh_ang} is fulfilled. On the contrary, in
Fig.~\ref{fig:henon3d_all_ang}b the distribution for $\theta_2$ is
not separated from the origin, i.e., the first and the second
subspaces are not hyperbolically isolated, so that the
case~\eqref{eq:henon3d_511} is not pseudohyperbolic.

\begin{figure}[tbp]
  \onefig{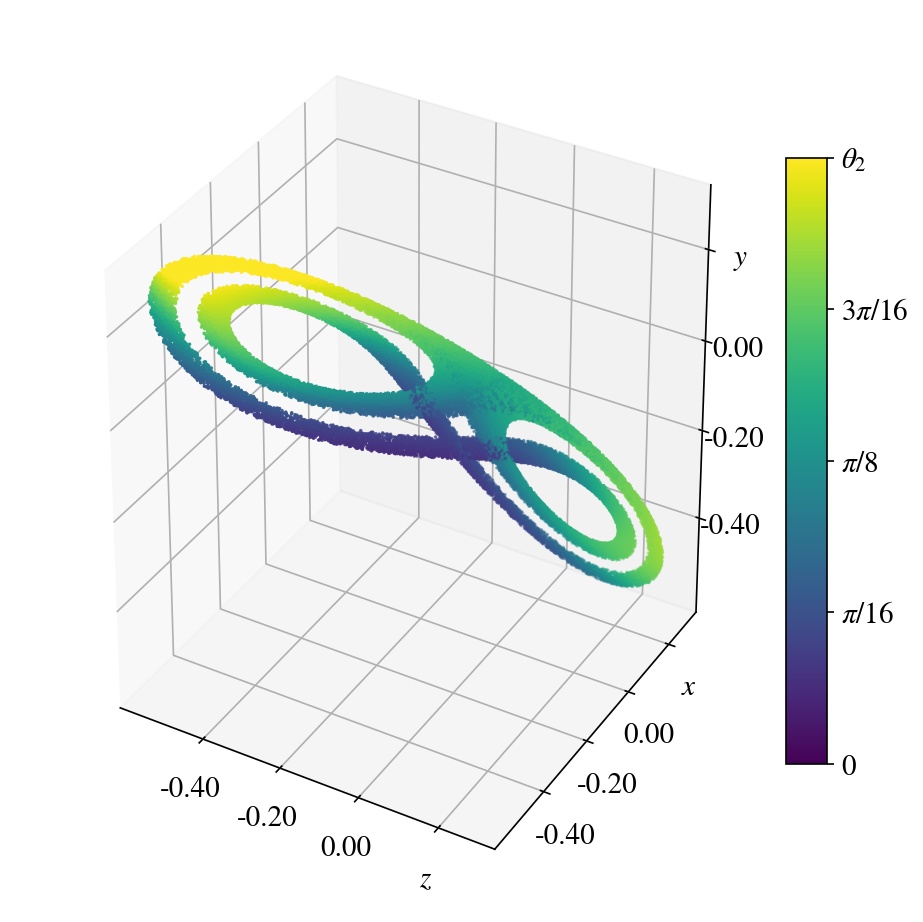}
  \caption{\label{fig:henon3d_52_attr3d}Attractor of the
    system~\eqref{eq:henon3d} with parameters~\eqref{eq:henon3d_52}.
    Colors represent values of $\theta_2$. Observe location of small
    angles in the middle area.}
\end{figure}

\begin{figure}[tbp]
  \onefig{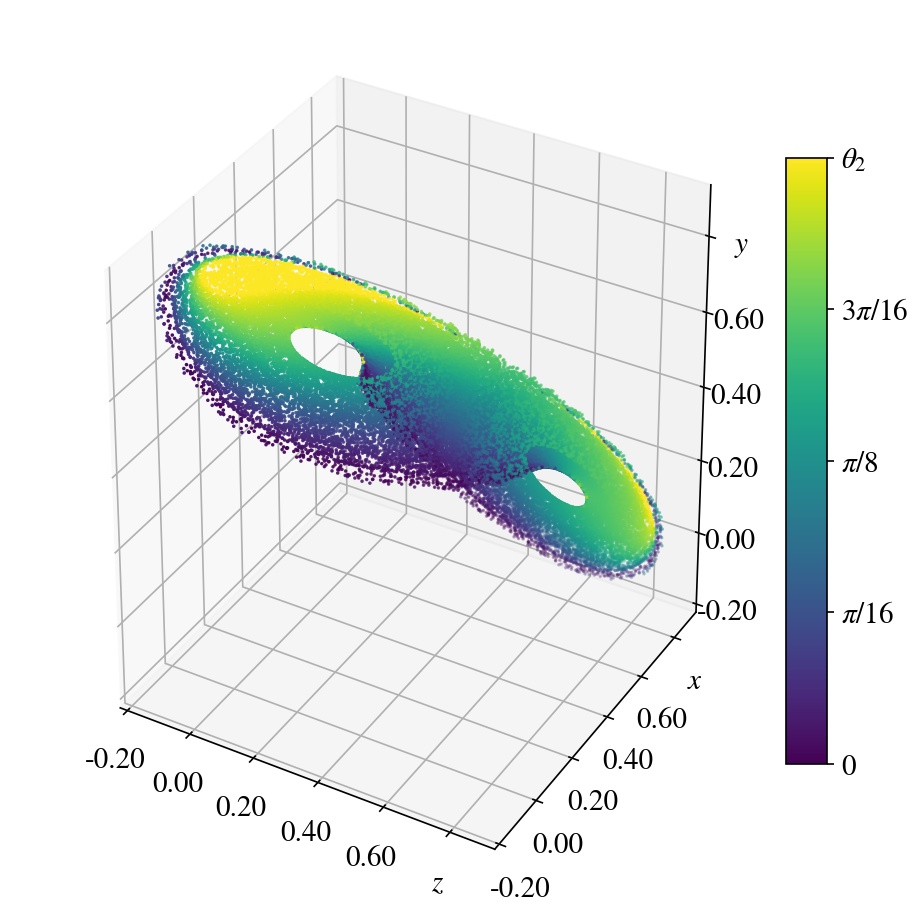}
  \caption{\label{fig:henon3d_511_attr3d}Attractor of the
    system~\eqref{eq:henon3d},~\eqref{eq:henon3d_511}. Observe
    vanishing angles on the edges.}
\end{figure}

\begin{figure}[tbp]
  \onefig{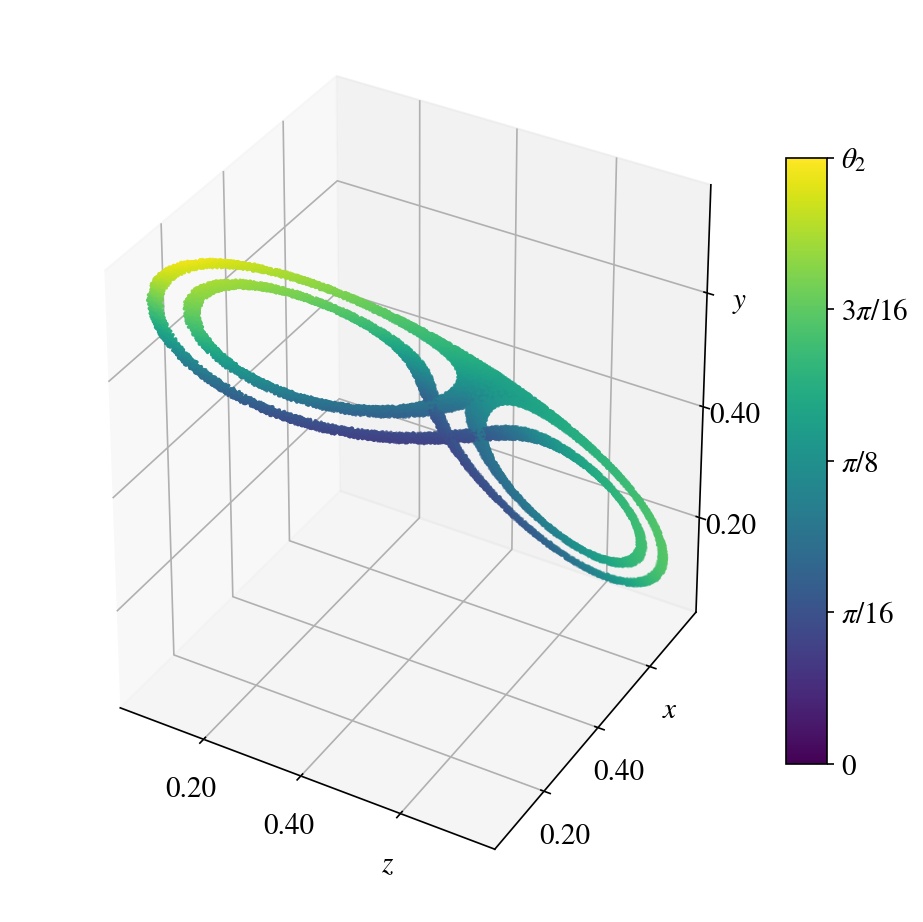}
  \caption{\label{fig:henon3d_512_attr3d}Attractor of the
    system~\eqref{eq:henon3d},~\eqref{eq:henon3d_512}. Observe
    similarity with the attractor in
    Fig.~\ref{fig:henon3d_52_attr3d}.}
\end{figure}

Phase portraits of the system~\eqref{eq:henon3d} with
parameters~\eqref{eq:henon3d_52},~\eqref{eq:henon3d_511},
and~\eqref{eq:henon3d_512} are shown in
Figs.~\ref{fig:henon3d_52_attr3d},~\ref{fig:henon3d_511_attr3d},
and~\ref{fig:henon3d_512_attr3d}, respectively. Colors represent
values of the angle $\theta_2$. Observe high similarity of the
pseudohyperbolic attractors in Figs.~\ref{fig:henon3d_52_attr3d}
and~\ref{fig:henon3d_512_attr3d}. Their small (but nonzero) angles
$\theta_2$ are located on bands crossing in the center of the
attractor.  On the non pseudohyperbolic attractor in
Fig.~\ref{fig:henon3d_511_attr3d} small and zero angles are located on
edge areas.

The instant exponents IBLEs and ICLEs are not applicable to discrete
time systems like~\eqref{eq:henon3d} since the local expansions and
contractions are explored by FTLEs and FTCLEs computed for one step of
time. Hence, we will consider only finite time exponents. Moreover,
the distributions of $\bar{S}_n$, $\ftcle_n$, and $\bar{D}_n$ will be
represented only for the parameters~\eqref{eq:henon3d_52} since two
other cases produce similar pictures.

\begin{figure}[tbp]
  \onefig{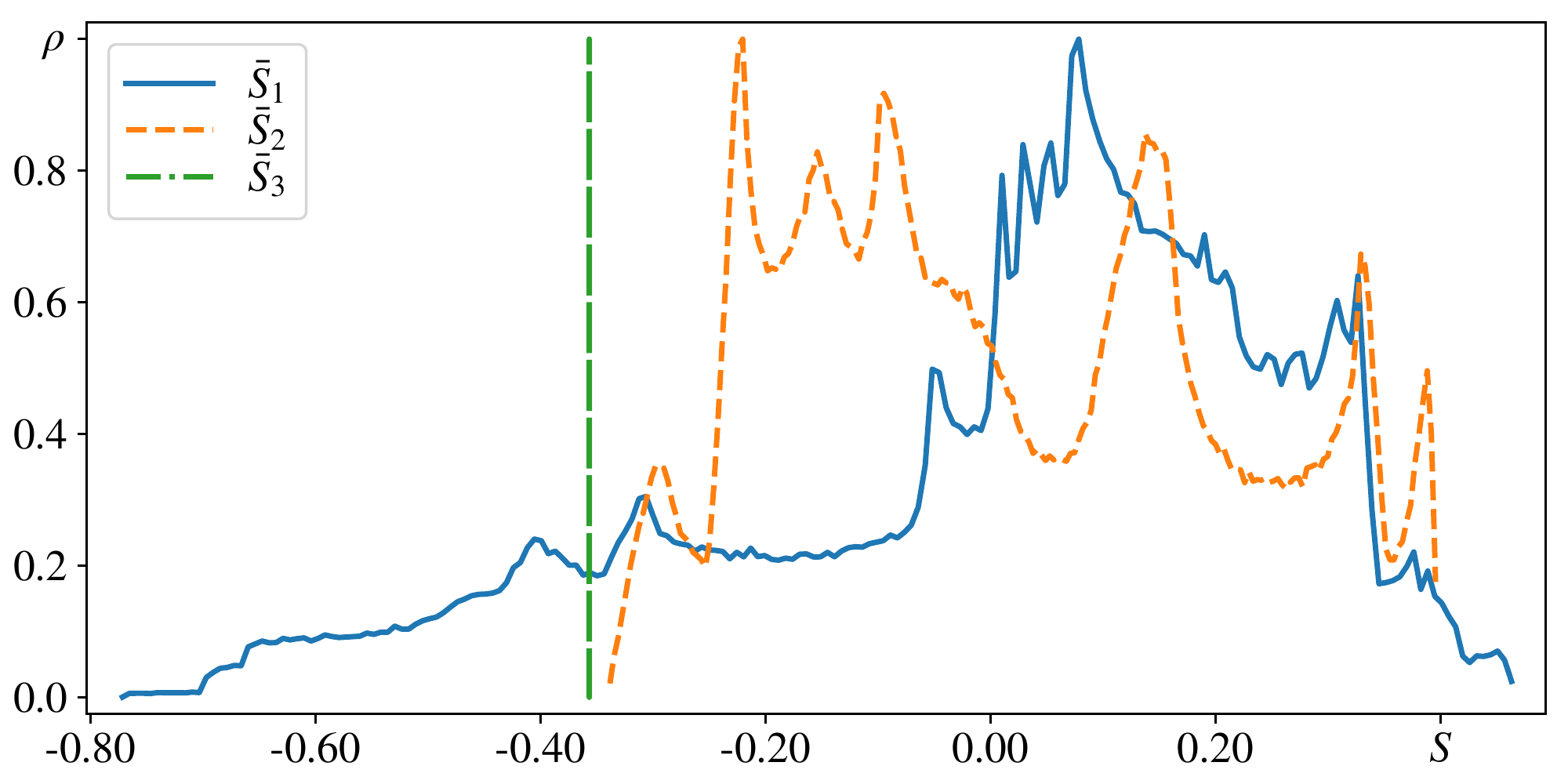}
  \caption{\label{fig:henon3d_52_acm}Distributions of summed FTLEs
    $\bar{S}_n(t,t+1)$ for the system~\eqref{eq:henon3d} with
    parameters \eqref{eq:henon3d_52}. Observe that all three
    fluctuating values can change signs.}
\end{figure}

\begin{figure}[tbp]
  \onefig{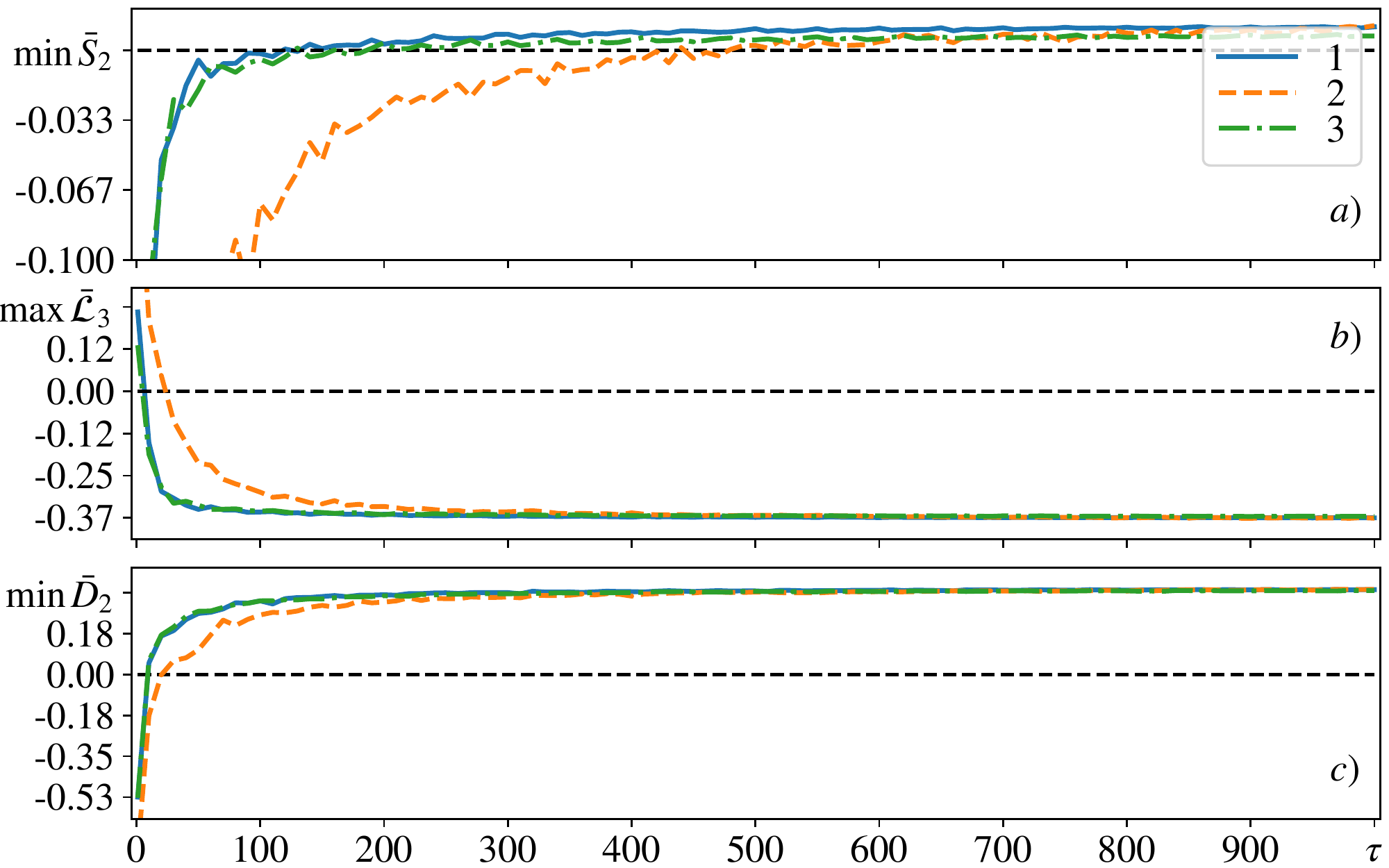}
  \caption{\label{fig:henon3d_all_minmax}As in
    Fig.~\ref{fig:lorenz_minmax} for the system~\eqref{eq:henon3d}
    with parameters~\eqref{eq:henon3d_52},~\eqref{eq:henon3d_511},
    and~\eqref{eq:henon3d_512}, respectively, curves 1, 2, and 3.
    Observe almost perfect coincidence of the curves 1 and 3
    corresponding to the pseudohyperbolic cases.}
\end{figure}

Figure~\ref{fig:henon3d_52_acm} shows that for the
case~\eqref{eq:henon3d_52} the property~\ref{it:psh_expand} is locally
violated and the first subspace is not strictly volume expanding. The
indication is that $\bar{S}_2$ oscillates being often positive and
negative. Analogously $\bar{S}_2$ oscillates for the
cases~\eqref{eq:henon3d_511} and \eqref{eq:henon3d_512} so that the
property~\ref{it:psh_expand} is also not fulfilled locally.  As one
can see in Fig.~\ref{fig:henon3d_all_minmax}a the lower boundary of
the distribution $\minval \bar{S}_2$ becomes positive only at
sufficiently large time scales in all three considered cases. Observe
almost identical behavior of $\minval \bar{S}_2$ for pseudohyperbolic
attractors, see curves 1 and 3 corresponding to
parameters~\eqref{eq:henon3d_52} and~\eqref{eq:henon3d_512},
respectively. For the non pseudohyperbolic attractor at
parameters~\eqref{eq:henon3d_511} the first subspace also becomes
strictly expanding, i.e., $\minval \bar{S}_2$ turns positive, but at
much higher time scale. As for the distribution for $\bar{S}_3$ in
Fig.~\ref{fig:henon3d_52_acm}, $\delta$ peak indicates that the
contraction in the whole tangent space of the
system~\eqref{eq:henon3d} is constant.

\begin{figure}[tbp]
  \onefig{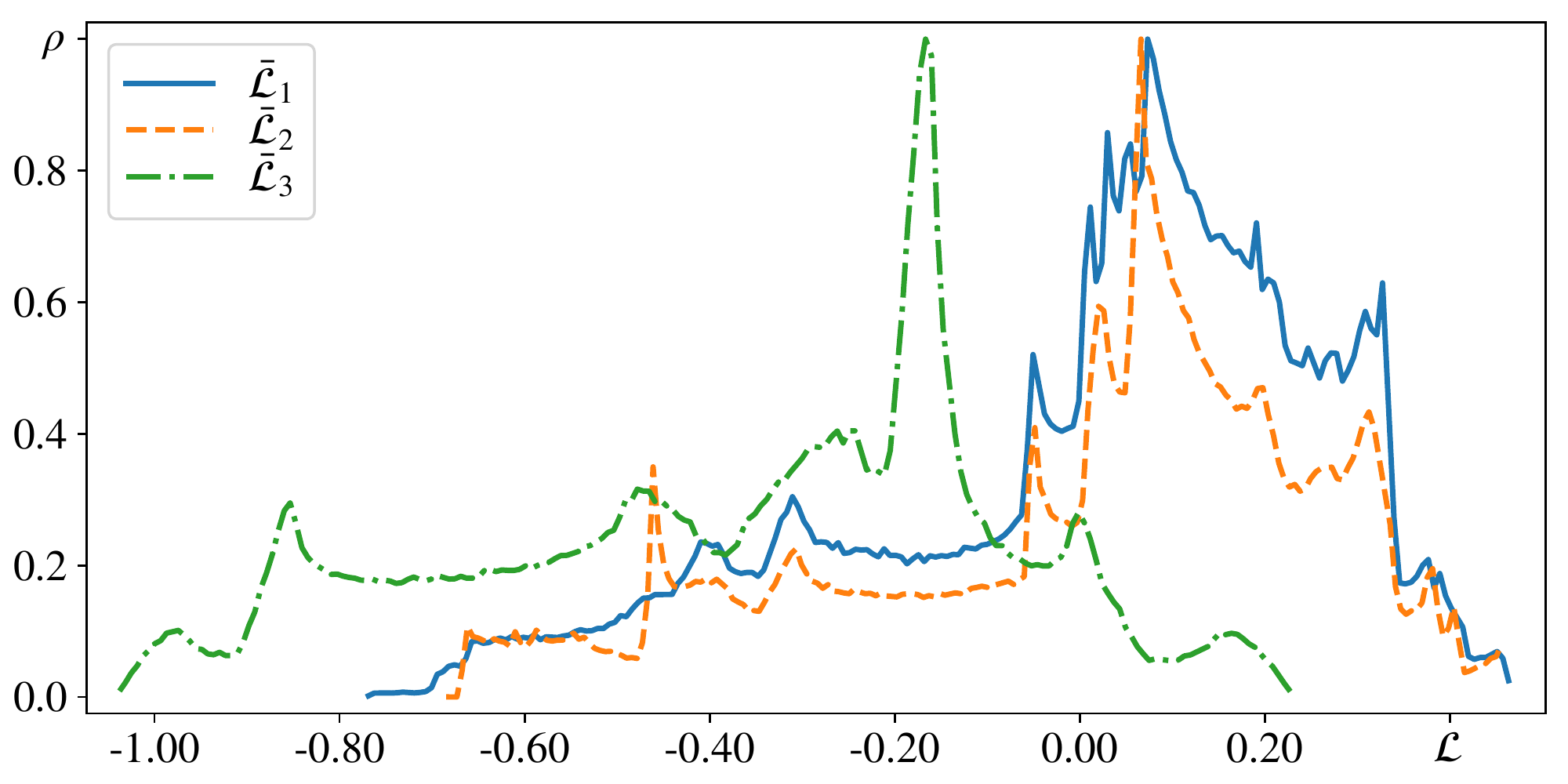}
  \caption{\label{fig:henon3d_52_icle}Distributions of FTCLEs
    $\ftcle_n(t,t+1)$ for the system~\eqref{eq:henon3d} with
    parameters \eqref{eq:henon3d_52}. Observe that all three FTCLE can
    be both positive and negative.}
\end{figure}

Figure~\ref{fig:henon3d_52_icle} demonstrates local violation of the
property~\ref{it:psh_contr} for the parameters~\eqref{eq:henon3d_52}:
$\ftcle_3$ responsible for the contraction in the second subspace can
sometimes be positive. Also $\ftcle_3$ demonstrates similar behavior
for parameter~\eqref{eq:henon3d_511} and~\eqref{eq:henon3d_512}.
Figure~\ref{fig:henon3d_all_minmax}b shows that the second subspace
for all three parameter sets becomes contracting when averaging time
$\ftledt$ grows. Again two pseudohyperbolic
cases~\eqref{eq:henon3d_52}, and~\eqref{eq:henon3d_512}, curves 1 and
3, respectively, behave almost identically, and the non
pseudohyperbolic attractor at parameters~\eqref{eq:henon3d_511}, curve
2, becomes contracting much later than two others.

\begin{figure}[tbp]
  \onefig{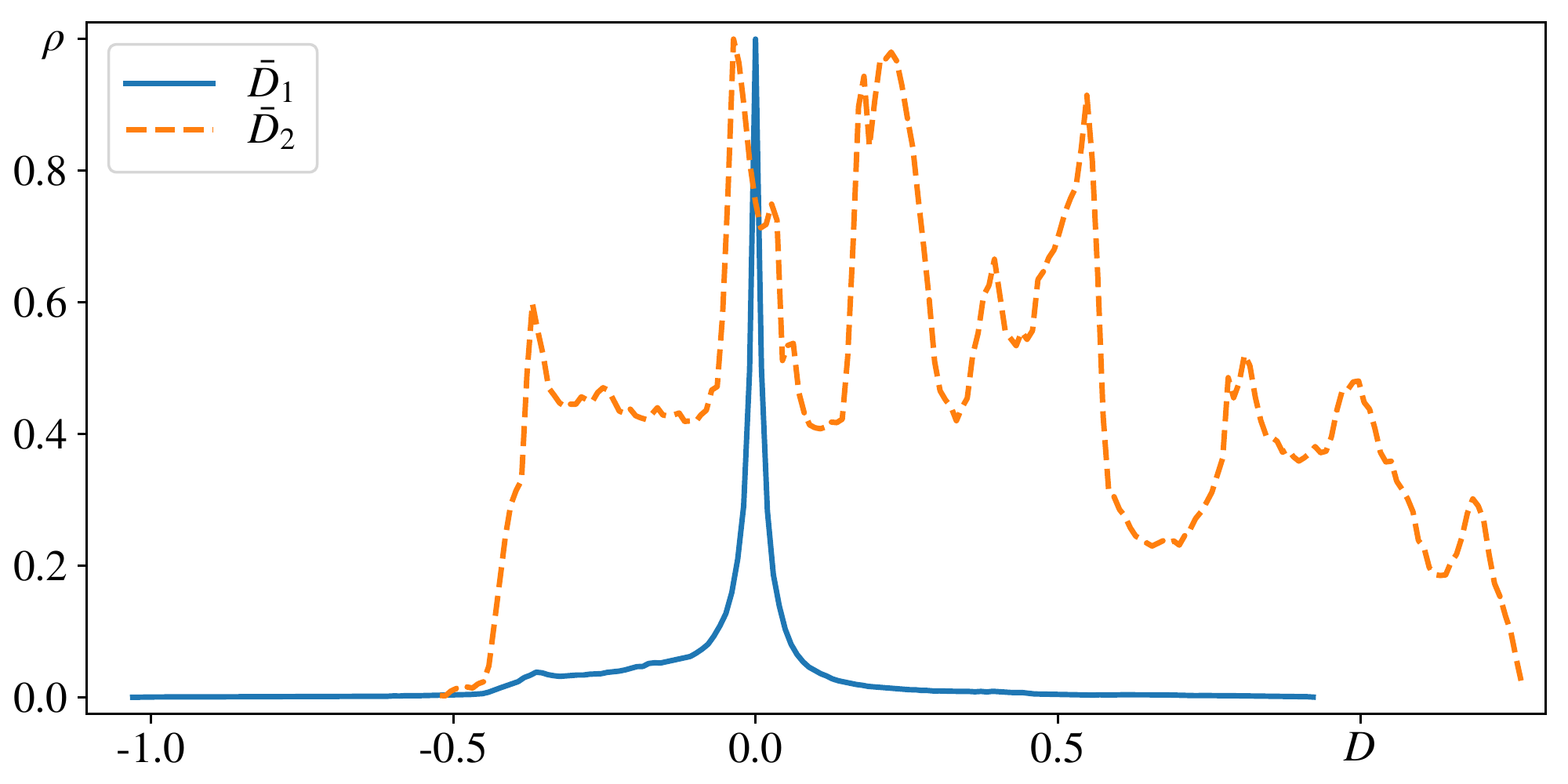}
  \caption{\label{fig:henon3d_dosmmx}Distributions of distances
    between FTCLEs $\bar{D}_n(t,t+1)$ for the
    system~\eqref{eq:henon3d} with parameters~\eqref{eq:henon3d_52}.
    Observe that two represented values oscillate around zero.}
\end{figure}

Finally, the property~\ref{it:psh_dosmmx} is also fulfilled only on
average, i.e., contraction in the first subspace can locally be
stronger than contraction in the second one. One can see in
Fig.~\ref{fig:henon3d_dosmmx} that indicating it $\bar{D}_2$ appears
both in positive and in negative semiaxes. Similarly $\bar{D}_2$
behaves for the cases~\eqref{eq:henon3d_511}
and~\eqref{eq:henon3d_512}. Only at approximately $\ftledt>10$
$\minval \bar{D}_2$ becomes positive for all three cases, see
Fig.~\ref{fig:henon3d_all_minmax}c. Again notice the coincidence of
the curves 1 and 3, representing pseudohyperbolic
cases~\eqref{eq:henon3d_52} and \eqref{eq:henon3d_512}, respectively.

Altogether, the pseudohyperbolicity of the system \eqref{eq:henon3d}
with parameters~\eqref{eq:henon3d_52} and \eqref{eq:henon3d_512} is
confirmed by the fulfillment of the property~\ref{it:psh_ang}, i.e.,
by the non vanishing angle between the first two-dimensional subspace
and the second one-dimensional one. The case~\eqref{eq:henon3d_511} in
agreement with above mentioned results is not pseudohyperbolic. The
three other properties~\ref{it:psh_expand},~\ref{it:psh_contr} and
\ref{it:psh_dosmmx} are violated locally. They are fulfilled only
after averaging on certain time scale.

\section{\label{sec:conl}CONCLUSION}

We have tested local structure of chaotic attractors related to
pseudohyperbolicity. Classical Lorenz system has been discussed as a
well known representative of pseudohyperbolic systems, and the R\"ossler
system is compared with it as an example of a system not belonging to
this category. Moreover several recently reported
examples~\cite{GonOvs05,GonGonPshyp,WildSpiral} of systems with and
without pseudohyperbolicity have been analyzed.

The main criterion of the pseudohyperbolicity is the splitting of the
tangent space into two hyperbolically isolated subspaces, volume
expanding and contracting ones. It means that the angles between these
two subspaces are nonzero at every point of the attractor. We have
computed numerically the corresponding angle distributions and
discussed the presence or absence of the pseudohyperbolicity in the
considered systems.

The properties of the two tangent subspaces of pseudohyperbolic
systems are usually explored via Lyapunov exponents $\lambda_i$. The
first $n$-dimensional subspace of a pseudohyperbolic system has to be
volume expanding so that $\sum_{i=1}^n\lambda_i>0$, and the second
subspace is contracting, i.e., $\lambda_i<0$ for $i>n$. Moreover, as
discussed in papers~\cite{TurShil98,TurShil08,PsReview,GonGonPshyp}, a
contraction if occurs in the first subspace, has to be weaker than any
contraction in the second subspace. However, Lyapunov exponents
describe attractors globally and the local properties are not taken
into account. Therefore, we have analyzed local, i.e., related to
infinitesimal and short time intervals volume expanding and
contracting properties of the two tangent subspaces.

To analyze expansion in tangent space on infinitesimal time we have
introduced a family of instant Lyapunov exponents. Unlike the well
known finite time ones, the instant Lyapunov exponents show expansion
or contraction on infinitesimal time intervals. Two types of instant
Lyapunov exponents are defined. One is related to ordinary finite time
Lyapunov exponents (FTLEs) computed in the course of standard
algorithm for Lyapunov exponents. These instant exponents are based on
orthogonal Gram-Schmidt vectors also known as backward Lyapunov
vectors and we refer to them as IBLE. Their sums reveal volume
expanding properties: the sum of the first $n$ IBLEs is the exponent
of growth or contraction of an $n$-dimensional tangent volume on
infinitesimal time. The other type of instant Lyapunov exponents shows
how covariant Lyapunov vectors grow or decay on infinitesimal time and
thus are called ICLE. They are appropriate for analysis of instant
single expanding or contraction direction in the tangent space.

Using both instant and finite time Lyapunov exponents, we have
demonstrated that for the Lorenz system the second subspace is
contracting on infinitesimal times and any instant contraction in the
first subspaces is always weaker than the contraction in the second
one. But the first subspace is not strictly volume expanding when
considered on infinitesimal times. This property turns fulfilled only
when the volumes evolution is observed on sufficiently large finite
time scales. For other tested systems all expanding and contracting
properties specific to the pseudohyperbolicity are observed only on
finite times. Instantly volumes from the first subspace can sometimes
be contracting, directions in the second subspace can sometimes be
expanded, and the instant contraction in the first subspace can
sometimes be stronger than the contraction in the second subspace.

\section*{ACKNOWLEDGMENTS}

Work of SPK on theoretical formulations was supported by grant of
Russian Science Foundation No 15-12-20035. The work of PVK on
elaborating computer routines and numerical computations was supported
by grant of RFBR No 16-02-00135.

\bibliography{pshyp}

\endpaper

\end{document}